\documentclass[12pt]{article}

\usepackage{microtype}
\usepackage{graphicx}
\usepackage{subfigure, setspace}
\usepackage{booktabs} % for professional tables
\usepackage{amsthm}
\usepackage{amsmath}
\usepackage[margin=1in]{geometry}
\usepackage{cite}
\usepackage[authoryear]{natbib}
\usepackage{graphicx}
\usepackage{amsfonts}
\usepackage{amssymb}
\usepackage{dsfont}
\usepackage{verbatim}
\usepackage{wrapfig}
\usepackage{upgreek}
\usepackage{setspace,enumitem}
\usepackage{float}
\usepackage{algorithm, algorithmic}
\usepackage{color,soul}
\usepackage[table,xcdraw]{xcolor}
\usepackage{adjustbox}
\usepackage{bbm}
\usepackage{multicol}
\usepackage{caption}

\newcommand\numberthis{\addtocounter{equation}{1}\tag{\theequation}}
\DeclareMathOperator{\E}{\mathbb{E}}

\theoremstyle{definition}
\newtheorem{definition}{Definition}[section]
\newtheorem{thm}{{\bf \sc Theorem}}[section]

\usepackage{hyperref}

\begin{document}

\title{Inference for Network Regression Models with Community Structure~\\~\\}

\author{Mengjie Pan\\Facebook \and Tyler H. McCormick\\University of Washington \and Bailey K. Fosdick\thanks{Contact information: Mengjie Pan (panmj0221@gmail.com), Tyler McCormick (tylermc@uw.edu), Bailey Fosdick (bailey.fosdick@colostate.edu)}\\Colorado State University}
\date{}

\maketitle
\begin{abstract}
Network regression models, where the outcome comprises the valued edge in a network and the predictors are actor or dyad-level covariates, are used extensively in the social and biological sciences. Valid inference relies on accurately modeling the residual dependencies among the relations.  Frequently homogeneity assumptions are placed on the errors which are commonly incorrect and ignore critical, natural clustering of the actors.  In this work, we present a novel regression modeling framework that models the errors as resulting from a community-based dependence structure and exploits the subsequent exchangeability properties of the error distribution to obtain parsimonious standard errors for regression parameters. 
\end{abstract}

\doublespacing 
\section{Introduction}
Researchers are often interested in how relations between pairs of actors are related to observable covariates, such as demographic, sociological and geographic factors. 
For example, \citet{ward2007persistent} examine political and institutional effects on international trade and find that the domestic political framework of the exporter and importer are important factors of the trade; \citet{aker2010information} explores the impact of mobile phones on the price difference of grain between a pair of markets and find that the introduction of mobile phone service explains a reduction in grain price dispersion; \citet{fafchamps2007formation} explore the role of geographic proximity on risk sharing among agriculture workers in the Philippines and find that intra-village mutual insurance links are largely determined by social and geographical proximity, potentially since personal/geographical closeness facilitates enforcement. 

In this work, we focus on a case where continuous relations between pairs of actors are modeled as a linear function of observable covariates. Continuous, pairwise relations can be represented as a network with directed, weighted edges.  We assume a set of observed covariates for the actors and dyads (ordered actor pairs) and wish to study the association between the relational response and covariates. Efficient inference for the effect of the covariates on the relations requires accurate modeling of the dependence between the regression errors. Our main contribution is a novel non-parametric block-exchangeability assumption on the covariance structure of the error vector suitable for when there is excess hidden block variation in the network beyond that accounted for by the covariates. When the underlying error structure of data satisfies the model assumptions, we show that our inference procedures have correct confidence interval coverage. We present algorithms to both estimate the latent block structure and estimate the corresponding standard errors of the regression coefficients. Our goal in the paper is to provide a new approach to model and estimate the dependence between residual relations, which bridges the gap between the existing non-parametric estimators.

Let $n$ be the observed number of individuals, $y_{ij}$ be the directed relational response from actor $i$ to actor $j$, and  $\boldsymbol{X}_{ij}= [ 1, X_{1,ij}, \cdot\cdot\cdot,  X_{(p-1),ij} ]^{T}$ be a $(p\times 1)$ vector of covariates. We assume there is no relation from an actor $i$ to itself. The regression model can be expressed
\begin{equation}
 y_{ij}=\boldsymbol{\beta}^T \boldsymbol{X}_{ij} + \xi_{ij}, \hspace{.1in}  i,j \in \{ 1,...,n\}, i \neq j.
 \label{eqn:linear_regression}
\end{equation}
We model the error vector ${\Xi}=[ {\xi}_{21}, {\xi}_{31}, \cdot\cdot\cdot, {\xi}_{n1}, \cdot\cdot\cdot, {\xi}_{1n}, \cdot\cdot\cdot, {\xi}_{(n-1)n} ]^T \sim N (\mathbf{0}, \Omega) $, where  $\Omega=\text{Var}(\Xi)$ is a $n(n-1)$ by $n(n-1)$ symmetric matrix. For example, if we are interested in how geographical and demographic factors affect number of mobile calls between actors,, then $y_{ij}$ is the number of mobile calls from actor $i$ to actor $j$ and $X_{ij}$ may include the actors' geographical distances and their mobile plans. Making inference on $\boldsymbol{\beta}$ then provides insights into how a change in geographical distance or mobile plans is associated with a change in the number of mobile calls.

In order to get accurate estimation of the standard error of $\boldsymbol{\beta}$ and thus a confidence interval with the correct coverage, we need to pose assumptions on the error structure that is satisfied by the data. The challenge in modeling $\Omega=\text{Var}(\Xi)$ is that $ \xi_{ij}$ and $ \xi_{kl}$ are likely correlated whenever the relation pairs share a member, i.e.$    \{i,j  \} \cap  \{k,l  \}  \neq \emptyset$ (\citet{kenny2006dyadic}). 
The residuals represent variation in relational observations not accounted by observable covariates, and two residuals which both involve actor A may be affected by actor A's individual effects. For example, if the residual of number of mobile calls from actor A to actor B is negative, we may expect number of mobile calls from actor A to actor C is likely also less than expected under the mean model, because actor A does not use mobile calls a lot. Another example is the case of reciprocal relations (\citet{miller1986reciprocity}). The residuals of number of mobile calls from actor A to actor B and from actor B to actor A are likely correlated, because they involve the same pair of actors and this dependence is often not fully captured by covariates. 

One set of approaches to model the covariance structure $\Omega$ is to impose parametric distributional assumptions on the error vector or model the error covariance structure directly (\citet{hoff2005bilinear}, \citet{ward2007persistent}, \citet{hoff2011separable}, \citet{hoff2015multilinear}). While these approaches produce interpretable representations of underlying residual structure, they always assume the error structure is consistent with an underlying parametric model. 

Another set of approaches to model the covariance structure, $\Omega$, is using non-parametric methods. However, existing approaches either make no distributional assumptions and estimate $\mathcal{O}(n^3)$ parameters (see dyadic clustering estimator in \citet{fafchamps2007formation}), or assume exchangeability of the error vector $\Xi$ and estimate five parameters (\citet{marrs2017standard}).  The former approach results in a standard error estimator for $\beta$ that is extremely flexible yet extremely variable, whereas the latter approach assumes all actors are identically distributed and results in a relatively restricted estimator. The former approach is appealing when a researcher does not have any information on the error structure and wants to allow for heterogeneity, while the later approach is appealing when a researcher is more confident the errors are exchangeable and thus can enjoy the simplicity of the error structure and a fixed, small number of covariance parameters. 
Nevertheless, there are likely cases where a researcher has some information about the error structure, but the errors are not exchangeable. This calls for an approach that bridges the gap between these two existing methods. 

We propose an alternative block-exchangeable standard error estimator that assumes that actors have block memberships and actors within the same block are exchangeable (i.e. have relations that are identically distributed). Heterogeneity based on unobserved variables are quite common in networks, and relational observations between actors in the same block may have different patterns than relations between actors in different blocks. The stochastic block model (\citet{holland1983stochastic}, \citet{snijders1997estimation}) and degree-corrected stochastic block model (\citet{karrer2011stochastic}) have been proposed to model connectivity between actors based on latent block memberships and actor degree heterogeneities. Spectral clustering algorithms (\citet{rohe2011spectral}, \citet{qin2013regularized}) have also been proposed to estimate the hidden block membership for these models. By imposing the exchangeability assumption on the error vector conditioned on block membership of the actors, we take into account possible block structure in the network residuals and allow for heterogeneity between blocks. Specifically, we propose an algorithm that estimates the covariance matrix $\widehat{\Omega}$ given the block memberships, as well as a second algorithm to estimate block memberships using spectral clustering. We present theoretical results proving the block-exchangeable estimator outperforms the exchangeable estimator when the errors are block-exchangeable. Critically, if the distribution of the covariates is dependent on block membership, we see a larger difference in standard errors from the block-exchangeable estimator compared to those from the exchangeable estimator.

\section{Previous Methodology}
\label{sec:previous}

In a linear regression model of form \eqref{eqn:linear_regression}, there are a number of ways to model $\Omega$. \citet{fafchamps2007formation} propose a maximally flexible model for $\Omega$ subject to the single condition that Cov$(\xi_{ij}\xi_{kl})  = 0   $ if dyads $(i,j)$ and $(k,l)$ do not share a member, i.e. $    \{i,j  \} \cap  \{k,l  \}  = \emptyset$. No additional structure is placed on the $\mathcal{O}(n^3)$ remaining covariance terms. This method is known as dyadic clustering, denoted here `DC', and we let $\Omega_{DC}$ denote the covariance matrix under the \citet{fafchamps2007formation} assumption. \citet{fafchamps2007formation} propose a simple way to estimate the elements in $\Omega_{DC}$: $\widehat{\text{Cov}}(\xi_{ij},\xi_{kl})=r_{ij} r_{kl}$, where $r_{ij}$ and $r_{kl}$ are the residuals of the corresponding relations. While the DC estimator is extremely flexible, the estimator $\widehat{\Omega}_{DC}$ contains $\mathcal{O}(n^3)$ parameters and each element is estimated by a single product of residuals. This makes the estimator highly variable, which consequently leads to highly variable $\boldsymbol{\beta}$ standard errors estimates. 

In order to ease the computational burden and decrease the variance of dyadic clustering estimator, \citet{marrs2017standard} propose an exchangeability assumption on the error vector and a simple moment-based estimator for the covariance parameters resulting in $\widehat{\Omega}_E$. The errors in a relational data model are jointly exchangeable if the probability distribution of the error vector is invariant under simultaneous permutation of the rows and columns. \citet{li2002unified} argue that data generated under the variance component model, which assumes that the observation can be decomposed additively into multiple actor-level components, and the Social Relation Model (\citet{warner1979new}, \citet{cockerham1977quadratic}) satisfy this exchangeability assumption. Under exchangeability and the assumption that the covariance between relations involving non-overlapping dyads is zero, \citet{marrs2017standard} shows that there are five non-zero parameters in $\Omega$ (see Figure \ref{fig:block_exch_covariance}), notably one variance $\sigma^2$ and four covariances $\{\sigma_A^2,\sigma^2_B,\sigma^2_C,\sigma_D^2\}$   They estimate these five parameters by averages of the corresponding residual products, greatly reducing the variance of the estimator $\widehat{\Omega}_E$ compared to $\widehat{\Omega}_{DC}$.

While the number of parameters is significantly reduced under the exchangeability assumption, this assumption may be violated in practice in many scientific settings. For example, when a network has block structure (i.e. community structure), such that actors in different blocks have different behavior patterns, this needs to be accounted for. For instance, conditioned on the covariates, relations in one block may have larger variation than those among actors in another block, thus violating the exchangeability assumption where a single variance is shared among all relations. In the mobile calls example, variance of phone calls among employed actors and that among unemployed actors may likely be different, meaning that without information on actor employment status, residual heterogeneity is likely present. This motivates us to consider a block-exchangeability assumption on $\Omega$.

\section{Block-exchangeability}
\label{sec:model}
With the dyadic cluster estimator making a single assumption but yielding too many parameters and the exchangeable estimator making strong assumptions, we propose a block-exchangeability assumption that compromises between imposing assumptions on error vector and model complexity. In a network of $B$ latent blocks, let $g_i$ denote the block assignment of actor $i$: $g_i \in \{1,...,B \}$. We propose the following definition of \textbf{block-exchangeability} as conditional exchangeability (\citet{lindley1981role}) of $\Xi$ given $g$: 
\begin{definition}
The errors in a relational data model are jointly block-exchangeable if $P(\Xi)$, the probability distribution of the error vector, is invariant under permutation of the rows and columns within each block:
\begin{center}
$P(\Xi)=P(\prod (\Xi))$ such that {$ g_i=g_{\pi(i)}$} and  {$g_j=g_{\pi(j)}$,}
\end{center}
where $\prod (\Xi)=\{ \xi_{\pi(i)\pi(j)} \}$ is the residual matrix with its rows and columns reordered according to permutation operator $\pi$. 
\end{definition}

A different exchangeable block assumption in the regression settings is discussed in \citet{mccullagh2005exchangeability}, where the distribution of observations is invariant under permutations that preserve the block-to-block relationship structure, i.e. permutations $\pi$ such that $B(i, j)=B(\pi(i),\pi(j)) \forall i, j $, where $B(i,j)=1$ if $g_i=g_j$, and $B(i,j)=0$ otherwise. There are two key differences between this assumption and that we propose. One is that block-exchangeability in \citet{mccullagh2005exchangeability} is on the observations, whereas we propose block-exchangeability on the errors. The other is that the permutation in \citet{mccullagh2005exchangeability} only requires that $B(i, j)=B(\pi(i),\pi(j))=0$, meaning observations that are in different blocks remain in different blocks after permutation.

Under our block-exchangeability assumption and conditional on block membership, the covariance between two arbitrary errors $\xi_{ij}$ and $\xi_{kl}$ takes one of the following six values depending on the block memberships $\{g_i,g_j,g_k,g_l\}$ and relationships among the indices $\{ i, j,k,l\}$:
\begin{tabular}{ll}
 Var$(\xi_{ij})= \sigma_{(g_i,g_j)}^2$  & Cov$(\xi_{ij}, \xi_{il})=\phi_{B_{ (g_i,  \{g_j,g_l \} )}}$ \\
 Cov$(\xi_{ij},\xi_{ji}) = \phi_{A_{ \{g_i,g_j  \}}}$ & Cov$(\xi_{ij}, \xi_{kj})=\phi_{C_{(g_j,  \{g_i,g_k  \} )}}$ \\
 Cov$(\xi_{ij}, \xi_{kl})=0$ & Cov$(\xi_{ij}, \xi_{ki})=\phi_{D_{(g_i,  g_j,g_k   )} }$  \\
\end{tabular}
where $ \{ \} $ denotes unordered set and $()$ denotes ordered set.   Note that this notation is an expansion of that introduced in \citet{marrs2017standard} such that all non-zero parameters are now indexed by node block memberships.

\begin{figure}[ht]
\begin{center}
\centerline{\includegraphics[width=\columnwidth]{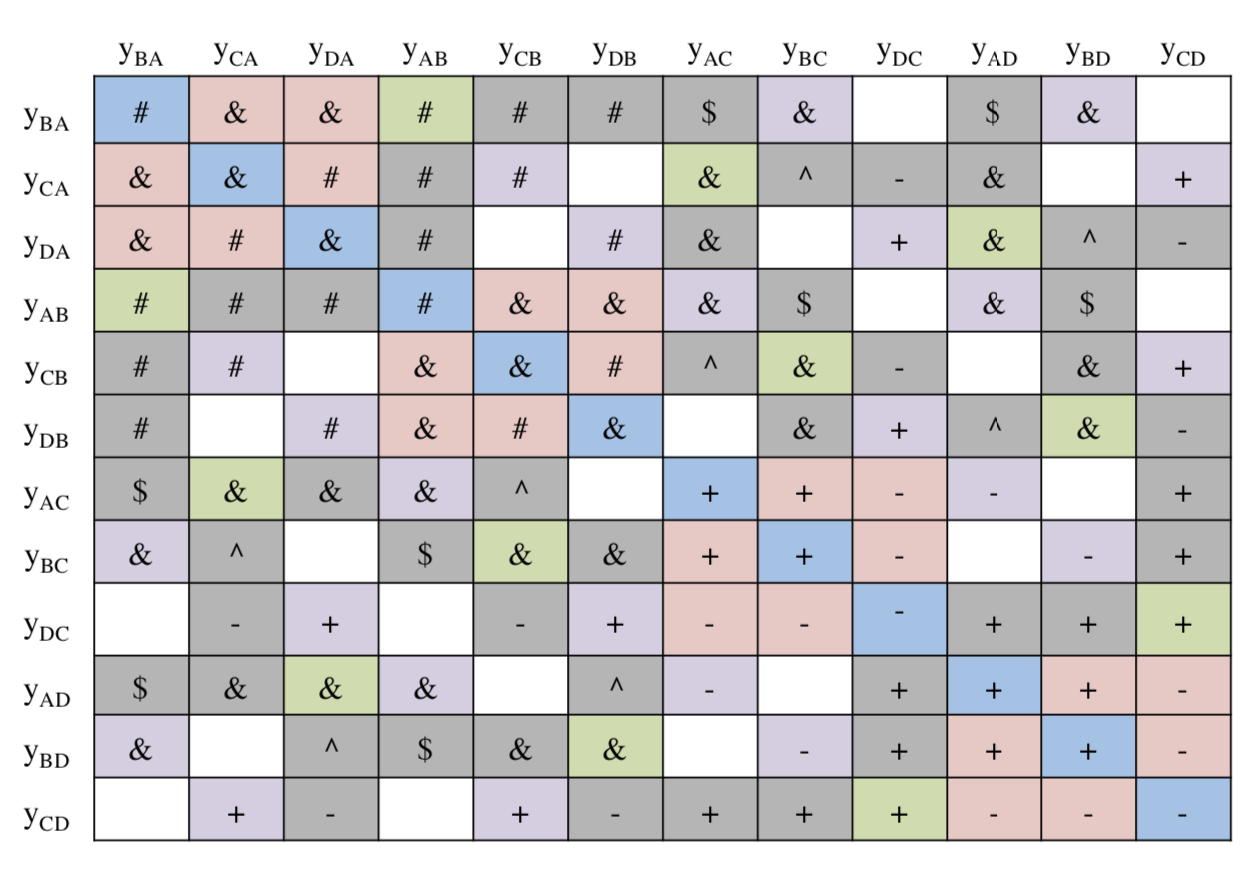}}
        \caption{Visualization of covariance matrix for a network of four actors. Under the block-exchangeability assumption, where $A$ and $B$ are in one block and $C$ and $D$ are in another block, entries shaded with the same color and symbol share the same parameter value. Conversely, under the exchangeability assumption, entries with the same color share the same value. } 
        \label{fig:block_exch_covariance}
\end{center}
\end{figure}

Figure \ref{fig:block_exch_covariance} shows a visualization of $\Omega_B$ for a simple network of four actors $\{A, B, C, D\}$, where actors A and B are in Block 1 and actors C and D are in Block 2. Under both exchangeability and block-exchangeability assumption, the blank entries indicate a covariance value of zero between non-overlapping dyads $(y_{ij},y_{kl})$ where $    \{i,j  \} \cap  \{k,l  \} = \emptyset$. Under the block-exchangeability assumption, each color denotes a dyad configuration and conditioned on the color, each symbol denotes a parameter indexed by the actor block memberships. Thus entries with the same color and symbol share the same parameter value. For example, $\text{Var}(\xi_{CA})=\text{Var}(\xi_{DA})=\text{Var}(\xi_{CB})=\text{Var}(\xi_{DB})=\sigma_{(2,1)}^2$, as denoted by the blue \& in Figure \ref{fig:block_exch_covariance},
because the sender is in Block 2 and the receiver is in Block 1. On the contrary, under the exchangeability assumption of \citet{marrs2017standard}, entries with the same color share the same value. Therefore, $\Omega_B$ has more parameters than $\Omega_E$, while maintaining the same places for zero-valued entries.

\begin{figure}[ht]
\begin{center}
\centerline{\includegraphics[width=\columnwidth]{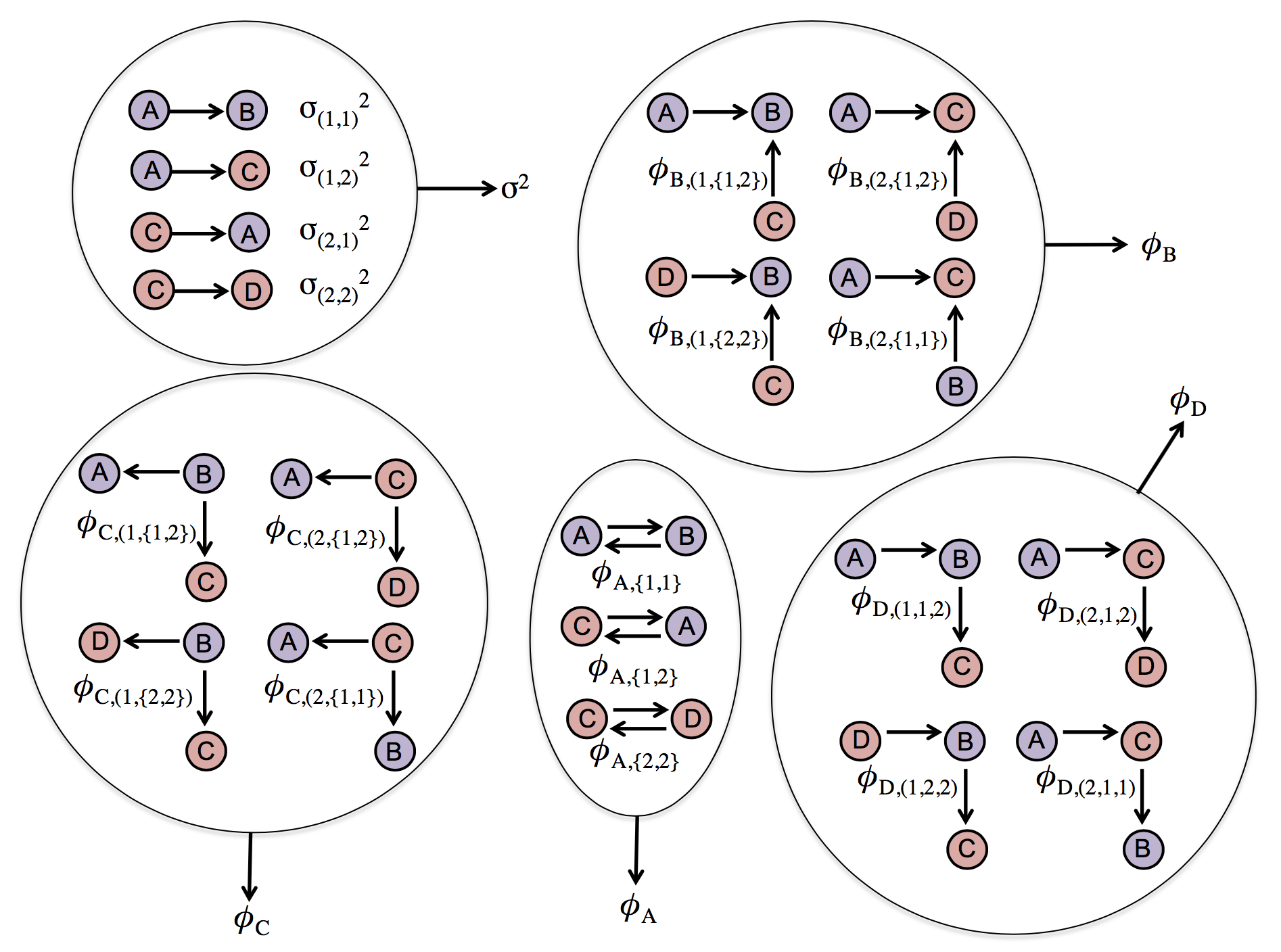}}
        \caption{Configurations of directed relation pairs under the block-exchangeability assumption in a simple network of four actors $\{$A, B, C, D$\}$, where A and B are in one block (indicated by purple color) and  C and D are in the other (indicated by light coral color). Each circle represents one dyad configuration, and the parameters within the circle correspond to those under the block-exchangeability assumption. The circles denote which block-exchangeable parameters share the same value under the exchangeability assumption and the associated exchangeable parameter is positioned at the end of the arrow outside the circle.}
        \label{fig:four_nodes_configuration}
\end{center}
\end{figure}

Figure \ref{fig:four_nodes_configuration} shows the  configurations of relation pairs under the block-exchangeability assumption. Each circle contains dyad configurations of the same type under exchangeability in $\Omega_E$.  However, under block exchangeability, there is variability with each configuration based on actor block memberships and these variations are shown within each circle. For example, the top left circle shows the variance parameters under block-exchangeability: $\sigma_{(1,1)}^2, \sigma_{(1,2)}^2, \sigma_{(2,1)}^2, \text{ and } \sigma_{(2,2)}^2$ corresponding to every ordered pair of blocks. In contrast,  
under the exchangeability assumption, all variance terms share the same parameter value $\sigma^2$. The top right circle shows four block-exchangeability parameters for the  configuration of relations pairs of the form $(y_{ij},y_{kj})$. In the top left corner of this circle, the common receiver actor B is in Block 1, sender A is in Block 1 and sender C is in Block 2 and therefore $\text{Cov}(\xi_{AB},\xi_{CB})=\phi_{B,(1,\{1,2\})}$. Because we only have two actors in each block, the case when $\text{Cov}(\xi_{ij},\xi_{kj})=\phi_{B,(1,\{1,1\})}$ is not shown in Figure \ref{fig:four_nodes_configuration} since it would require three actors $i, j, \text{ and } k $ in Block 1. In general the number of block-exchangeable parameters belonging to each configuration type depends on the number of blocks $B$ (see Table \ref{table:num_parameters}).

\begin{table}
\centering
\begin{tabular}{ c c }
\hline 
Covariance term & Number of parameters\\
\hline
Var$(\xi_{ij})=\sigma^2$ &$B^2$  \\ 
Cov$(\xi_{ij}, \xi_{ji})=\phi_A$ & $ B(B+1)/2$ \\  
Cov$(\xi_{ij}, \xi_{il})=\phi_B$ & $B^2(B+1)/2$ \\
Cov$(\xi_{ij}, \xi_{kj})=\phi_C$ &  $B^2(B+1)/2$\\
Cov$(\xi_{ij}, \xi_{ki})=\phi_D$ & $B^3$\\
 \hline
\end{tabular}
\caption{Number of parameters of each covariance type under the block-exchangeability assumption, where $B$ is the number of actor blocks.  Note that if there are fewer than three actors in a given block, some of these parameters will not appear in $\Omega_B$.  }
\label{table:num_parameters}
\end{table}

As shown in Table \ref{table:num_parameters}, the number of parameters in $\Omega_B$ is on the order of $\mathcal{O}(B^3)$. This is substantially greater than the number of parameters under the exchangeability assumption, which is five regardless of network size, yet significantly smaller than the number of parameters for dyadic clustering, which is on the order of $\mathcal{O}(n^3)$. The block-exchangeablility assumption balances between imposing assumptions on error vector and model complexity, in an attempt to model the covariance matrix with a reasonable number of parameters while keeping the assumptions feasible for real world applications.  

\section{Network Regression with Block-exchangeable Errors}
\label{sec:algorithm}
Assuming the errors are block-exchangeable and there are $B$ blocks, we now present algorithms that produce standard error estimates for the coefficients $\boldsymbol{\beta}$ in a linear regression model \eqref{eqn:linear_regression}.
Let $\boldsymbol{X}$ be the design matrix, $\widehat{\boldsymbol{\beta}} = (\boldsymbol{X}^T\boldsymbol{X})^{-1}\boldsymbol{X}^T y$ denote the ordinary least squares estimate of $\boldsymbol{\beta}$ and  $r_{ij}=y_{ij}-\boldsymbol{X} \widehat{\boldsymbol{\beta}}$ denote the residual for observation $y_{ij}$.  

\subsection{Known Blocks}
\label{sec:alg_known_blocks}
Given block memberships, the estimate of each block-exchangeable parameter is formed by the empirical average of the products of the residual pairs of the same block-dyad configuration type. To formally describe the estimator, let $[B] = \{1,...,B\}$ and $[n] = \{1,...,n\}$. Let M index the five dyad configurations $M \in \{\sigma^2, \phi_A, \phi_B,\phi_C,\phi_D\}$, and $Q_M$ denote the set of block pairs/triplets for dyad configuration $M$ given $[B]$.  Thus $Q_{\sigma^2}=\{  (u,v): u, v \in[B] \}$, and $Q_{\phi_B}=\{  (u,\{v,w \}): u, v, w \in [B] \}$.  We explicitly define all other sets $Q_M$ in the supplementary material.  Furthermore, let $\Phi_{M,q}$, where $q \in Q_M$, denote the set of ordered relation pairs that have the configuration $M$ and block specification $q$. Thus $\Phi_{\sigma^2,(u,v)}= \{ [(i,j),(i,j)]: i, j\in [n], i \neq j, {g}_i=u,   {g}_j=v\} $.  All other sets $\Phi_{\phi_A,\{u,v\}}$, $\Phi_{\phi_B,(u,\{v,w\})}$, $\Phi_{\phi_C,(u,\{v,w\})}$ and $\Phi_{\phi_D,(u,v,w)}$ are explicitly defined in the supplementary material.  Algorithm \ref{alg:known_blocks} formally describes estimation on ${\Omega}_{B}$ in this known block setting.

\begin{algorithm}[tb]
   \caption{Known block estimation of ${\Omega}_{B}$ }
   \label{alg:known_blocks}
\begin{algorithmic}
\STATE {\bfseries Input:}
residuals $\{r_{ij}:  i \neq j \}$, number of blocks $B$, block memberships $\{g_i\}$
   \STATE {\bfseries Output:}
$\widehat{\Omega}_B$ \\
   \STATE  
   1. For each configuration type $M$ and block combination $q$, calculate the set of residual products associated with $\Phi_{M,q}$:
\begin{center}
$\boldsymbol{R}_{M,q}= \{ r_{jk}r_{mn}: [(j,k),(m,n)] \in \Phi_{M,q} \}$
\end{center}
\STATE 3. Estimate the parameters in ${\Omega}_B$ using the empirical average of the corresponding residual products:
    $$\hat{\theta}_{M,q}=\frac{ \sum_{t: t\in \boldsymbol{R}_{M,q}}t}{ |\boldsymbol{R}_{M,q}|}$$
    where $\theta_{M,q}$ is the block-exchangeable parameter  corresponding to $M$ and $q$. 
\end{algorithmic}
\end{algorithm}

\subsection{Unknown Blocks}
\label{sec:alg_unknown_blocks}

When block memberships are unknown, we propose spectral clustering to estimate them by constructing a similarity matrix from the regression residuals (see Algorithm \ref{alg:unknown_blocks}). For each actor $i$ and each dyad configuration $M$, we extract all pairs of relation residuals that involve actor $i$ as the overlapping actor in the given configuration. Let $\Phi_{M,i}$ denote the set of relation pairs that involve a specific actor $i$ in configuration type $M \in \{\sigma^2, \phi_A, \phi_B,\phi_C,\phi_D\}$.  For example, $\Phi_{\sigma^2,i}= \{ [(i,j),(i,j)]: j\in [n], i \neq j \} \cup  \{ [(j,i),(j,i)]: j\in[n], i \neq j \}$ and  $\Phi_{\phi_B,i}= \{ [(i,j),(i,k)]: j,k \in [n], i \neq j \neq k \}$. Complete definitions of all other sets are provided in the supplementary material. We compute the Kolmogorov-Smirnov statistic between the distribution of residual products that involve actor $i$ and the distribution that involve actor $j$ for each configuration type $M$ and combine these to create a similarity measure between actors $i$ and $j$.  Unnormalized spectral clustering is then performed on the resulting similarity matrix to obtain block membership estimates (\citet{von2007tutorial}).

\begin{algorithm}[tb]
   \caption{Block membership estimation}
   \label{alg:unknown_blocks}
\begin{algorithmic}
\STATE {\bfseries Input:}
residuals $\{r_{ij}: i \neq j\}$, number of blocks $B$, $K$ for nearest neighbor graph
   \STATE {\bfseries Output:} estimated block memberships $\{\hat{g}_i\}$\\
 \STATE   
 1. For each actor $i$ and configuration $M \in \{\sigma^2, \phi_A, \phi_B,\phi_C,\phi_D\}$, calculate the set of residual products for $\Phi_{M,i}$:
\begin{center}
$\boldsymbol{R}_{M,i}= \{ r_{ab}r_{cd}: [(a,b),(c,d)] \in \Phi_{M,i} \}$ 
\end{center}
\STATE   
2. Let $F_{i,M}$ be the empirical distribution function for $\boldsymbol{R}_{M,i}$. For each pair of actors $i$ and $j$ and for each $M$, calculate the Kolmogorov-Smirnov statistic 
\begin{center}
$KS_{i,j,M}=\sup_x\limits |F_{i,M} (x) - F_{j,M} (x)|$. 
\end{center}
\STATE 3.    For each pair of actors $i$ and $j$, define 
\begin{center}
$s_{ij}=1-\left(\sum\limits_{M \in \{\sigma^2, \phi_A, \phi_B,\phi_C,\phi_D\}}\limits KS_{i,j,M} \right)/5$.
\end{center}
\STATE  4. Let $W=(w_{ij})_{i,j=1,...,n}$ denote the weighted adjacency matrix,where 
\[
w_{ij}=w_{ji}=
\begin{cases}
    s_{ij},& \text{if  } i \in KNN(j) \text{ or } j \in KNN(i) \\
    0,              & \text{otherwise}
\end{cases}
\]
where $KNN$ denotes K-nearest neighbor.
\STATE  5. Perform unnormalized spectral clustering  on weighted graph $W$ to get estimated blocks $\hat{g}_i$, where $\hat{g}_i \in [B] \;\; \forall i$.
\end{algorithmic}
\end{algorithm}

When block memberships are known, we apply Algorithm 1 to obtain $\widehat{\Omega}_B$. When block membership are unknown, we apply Algorithm 2 to estimate the block memberships $\{\hat{g}_i\}$ and then apply Algorithm 1 using $\{\hat{g}_i\}$ as an input to obtain $\widehat{\Omega}_B$. The value $K$ in step 4 of Algorithm 2 is a tuning parameter and is used to construct a K-nearest neighbor weighted adjacency matrix for input to the spectral clustering. \citet{maier2007cluster} prove that choosing $K=c_1 n - c_2\log(n) + c_3$, where $c_1, c_2 \geq 0$ and $c_3$ are all constants, provides an optimal choice of $K$. We found that for our simulation setting $K=0.2n$ worked well. When block memberships are known, computation of $\widehat{\Omega}_B$ is quite inexpensive because the algorithm simply extracts all dyad pairs with the same covariance and averages the residual products (e.g. 5 seconds when $n=80$,  20 seconds when $n=160$ on standard machine). When the block memberships are unknown, Step 1 and 2 of Algorithm 2 may be expensive if the network size is large. In these cases, we suggest a modification of Step 2. Instead of letting $F_{i,M}$ be the empirical distribution function for $\boldsymbol{R}_{M,i}$, we modify $F_{i,M}$ to be the empirical distribution function for quantiles of $\boldsymbol{R}_{M,i}$. This reduces the size of the set $\boldsymbol{R}_{M,i}$, which decreases the storage cost as well as the computational cost of computing Kolmogorov-Smirnov statistic.

\section{Theoretical Analysis of Estimator}
\label{sec:theory}
 If the block-exchangeability assumption is appropriate, then our method provides accurate estimation of the regression coefficient standard errors, and confidence intervals constructed with such standard errors have the correct coverage. This is why an accurate estimation of standard errors is important in inference on the coefficients.
 
Given $\widehat{\Omega}_B$ from Algorithm 1, the sandwich covariance estimator can be used to estimate the standard error of the ordinary least squares estimate $\widehat{\boldsymbol{\beta}}$: 
\begin{equation}
\widehat{V}(\widehat{\boldsymbol{\beta}})= (\boldsymbol{X}^T\boldsymbol{X})^{-1} \boldsymbol{X}^T \widehat{\Omega}_B \boldsymbol{X}(\boldsymbol{X}^T\boldsymbol{X})^{-1}.   
\label{eq:sandwich}
\end{equation}
Observe that entries in $\widehat{V}(\widehat{\boldsymbol{\beta}})$ are entries in $\widehat{\Omega}_B$ weighted by functions of $\boldsymbol{X}$. It is possible that even with the (incorrect) exchangeable covariance structure in $\Omega$, we still obtain accurate standard error estimation of $\widehat{\boldsymbol{\beta}}$ because the difference $\widehat{\Omega}_B - \widehat{\Omega}_E$ averages out over $\boldsymbol{X}$. Here we quantify the difference between the standard error estimator of $\widehat{\boldsymbol{\beta}}$ under the assumption of exchangeability and block-exchangeability, as a function of covariates $\boldsymbol{X}$, block assignments $\{g_i\}$, and the true $\Omega$. 

Consider a simple linear regression model with only one covariate:
\begin{equation}
y_{ij}=\beta_0+\beta_1 X_{ij}+\xi_{ij},  
\label{eq:simple-lm}
\end{equation}
where $y_{ij}$ is the observed relation, $X_{ij}$ is a scalar covariate, $\xi_{ij}$ is the error term, $\beta_0$ is the intercept, and $\beta_1$ is the covariate coefficient. In addition, assume there is a two block structure in the network, with block sizes $n_1$ and $n_2$, respectively, where $n_1 + n_2 = n$.  Under the assumption that the error vector is block-exchangeable, we show that the difference in $\widehat{V}(\widehat{\boldsymbol{\beta}})$ with the exchangeable estimator $\widehat{\Omega}_E$, denoted $\widehat{V}_E(\widehat{\boldsymbol{\beta}})$, and that with the block-exchangeable estimator $\widehat{\Omega}_B$, denoted $\widehat{V}_B(\widehat{\boldsymbol{\beta}})$, converges in probability to a matrix that depends on the distribution of the covariate, block assignments, and parameters in $\Omega_B$.

\begin{thm}
\label{thm:main-se}
Assume (a) the error vector satisfies the block-exchangeability assumption, with two blocks of sizes $n_1$ and $n_2$, (b) $\boldsymbol{X}$ is a full rank $(n(n-1) \times 2)$ matrix, (c) covariates $\{X_{ij}\}$ are independent and identically distributed, (d) the fourth moment of the errors and covariates are bounded, (e) errors $\Xi$ and $\boldsymbol{X}$ are independent, and (f) the number of blocks $B$ is $\mathcal{O}(1)$.
As $n_1 \rightarrow \infty,n_2 \rightarrow \infty $, and $ n_1/n_2 \rightarrow \alpha$, where $\alpha$ is a constant such that $0 < \alpha < \infty$,
    \begin{equation}
n \left( \widehat{V}_B(\widehat{\boldsymbol{\beta}})- \widehat{V}_E(\widehat{\boldsymbol{\beta}}) \right)  \overset{p}{\to} c (\boldsymbol{X}).
 \end{equation}
 where $c(\boldsymbol{X})$ is a weighted linear combination of the differences between the true block exchangeable parameters and the corresponding exchangeable parameters (when the block exchangeable parameters are appropriately averaged within configuration type) and convergence is pointwise. Furthermore, when $X_{ij}$ is independent of $g_i$ and $g_j$, $c(\boldsymbol{X})$ = $\boldsymbol{0}$ and thus the estimators are asymptotically equivalent.
\end{thm}

Proof of this theorem is provided in the supplementary materials. The corresponding exchangeable parameter $\sigma^2$ under block-exchangeability is a weighted average of $\sigma^2_{(1,1)},\sigma^2_{(1,2)}, \sigma^2_{(2,1)}$, and $\sigma^2_{(2,2)}$.  Note we can interpret $\sigma^2$ as the common variance term if the error vector is in fact exchangeable. We use the same logic for the other four configurations, and recognize that the magnitude and sign of the difference in standard errors using block-exchangeable estimator and exchangeable estimator are determined by a sum of weighted differences of all five configurations. Therefore, whether the exchangeable estimator has over- or under- coverage depends on parameters in $\Omega_B$, $\{g_i\}$, and $\{X_{ij}\}$.  

The second part of the theorem notes that even if the differences ${\sigma}_{(1,1)}^2-{\sigma}^2, {\sigma}_{(1,2)}^2-{\sigma}^2, {\sigma}_{(2,1)}^2-{\sigma}^2, {\sigma}_{(2,2)}^2-{\sigma}^2$ are nonzero, as long as the covariate $X_{ij}$ is independent of block memberships $g_i$ and $g_j$, on average the difference will disappear after adjusted by weights. This is a critical insight, because we see that in order for the block-exchangeable estimator to have lower bias than exchangeable estimator, we need (1) the error vector satisfies block exchangeability but not exchangeability, and (2) the distribution of $X_{ij}$ is correlated with on $g_i$ and $g_j$.

\section{Simulations}
\label{sec:simulation}
\begin{figure}[h!]
\begin{center}
\centerline{\includegraphics[width=1.1\columnwidth]{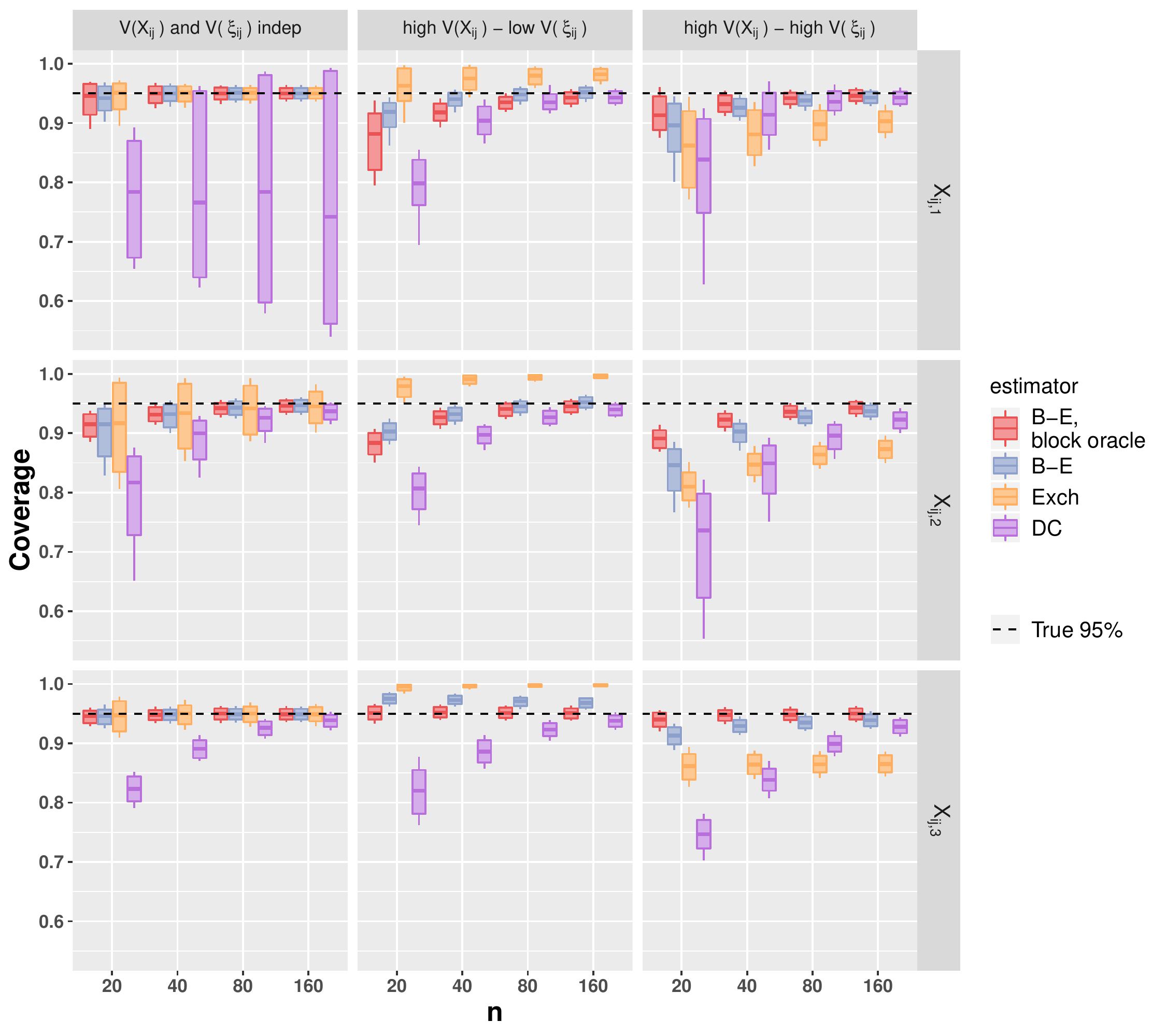}}
    \caption{Coverage of $95\%$ confidence interval for $\beta_1$ under three error settings and three covariate types for the block-exchangeable standard error estimator conditioned on the true block memberships (oracle), block-exchangeable standard error estimator with estimated blocks, exchangeable standard error estimator, and dyadic clustering standard error estimator.}
    \label{fig:main_sim_plots}
\end{center}
\end{figure}

To evaluate the performance of our proposed block-exchangeable error model, we generate data from a modified latent space model \citep{hoff2005bilinear}, which satisfies the requirements for block exchangeability.  We consider a simple regression model with one covariate, as in \eqref{eq:simple-lm} where both coefficients equal 1. We consider three settings for the relationship between the covariate and block structure, and three types of covariates.  Figure \ref{fig:main_sim_plots} shows the coverage of $95\%$ confidence intervals for $\beta_1$ for all nine simulation settings. The first column represents the cases where the covariate $X_{ij}$ is uncorrelated with block membership, the second column represents the case where relations with high variance in $X_{ij}$ are correlated with low variance errors $\xi_{ij}$, and the third column represents the case where relations with high variance in $X_{ij}$ also have  high variance errors $\xi_{ij}$. The rows represent different covariates: the first row is a binary indicator of actors sharing an attribute $X_{ij,1}=\mathds{1}_{[X_i=X_j]} $, the second row represents the absolute difference between an actor attribute $X_{ij,2}=|X_i-X_j| $, and the third row represents a pairwise covariate with block structure $X_{ij,3} \sim N(0,a_{g_i,g_j}^2)$. We generated 1000 errors for each of 500 simulations of the covariates and block memberships, and considered networks of size $20, 40, 80, \text{ and } 160$. We consider four estimators of $\Omega$ that are then plugged into the sandwich estimator \eqref{eq:sandwich} to obtain a confidence interval for $\beta_1$. The red box shows the coverage using the block-exchangeable estimator conditioned on the true block membership (Algorithm \ref{alg:known_blocks}), the blue box shows the coverage using the block-exchangeable estimator with the estimated block membership (Algorithms \ref{alg:known_blocks} and \ref{alg:unknown_blocks}), the yellow box shows the coverage using exchangeable estimator, and the purple box shows the coverage using the dyad clustering estimator. For each boxplot, the middle line indicates the median coverage, the top and bottom boundaries indicate the $90\%$ and $10\%$ percentiles, and the top and bottom whiskers indicate the $97.5\%$ and $2.5\%$ percentiles. 

The block-exchangeable estimator performs similarly to the exchangeable estimator when the covariate is uncorrelated the errors, while the block-exchangeable estimator substantially outperforms the exchangeable estimator when the covariate is correlated the errors. This is consistent with our theoretical results in Section \ref{sec:theory}. When high variance in a relation's covariate $X_{ij}$ is associated with low variance in the error $\xi_{ij}$, we observe that the exchangeable estimator is conservative, and the bias in coverage probability increases with increasing network size. On the contrary, the coverage bias of block-exchangeable estimator decreases with increasing network size. When high variance in a relation's covariate $X_{ij}$ is associated with high variance in the error $\xi_{ij}$, the exchangeable estimator is anti-conservative, and its performance improves little with increasing network size. Most notably, at $n=160$, the exchangeable estimator's coverage is worse than the dyadic clustering estimator, which is evidence that estimators with strict assumptions perform worse than distribution-free estimators when the assumptions are violated.  In addition, we observe that the differences between the oracle block-estimator using true block memberships and the block-estimator using estimated block membership decreases with increasing network size, suggesting that our block estimation gets better with increasing network size.

\section{Air Traffic Data}
\label{sec:dataanalysis}
We demonstrate our method on data representing passenger volume between US airports \citep{airport2016passenger}.  
The data consist of origin, destination, and number of passengers by month for $n=573$ airports for all months of 2016. The number of passenger seats is a right-tailed skewed distribution, so we use the values $y_{ij}^{'}=\log(y_{ij}+1)$ as the relational observations for regression model in \eqref{eqn:linear_regression}. For covariates, we calculated the great circle distance between two airports using their longitudes and latitudes. Additionally, we identified the county of the municipality of each airport, and found the total GDP of that county from \citet{GDP2015} and average payroll of an employed person from \citet{industry2015}. We standardized the distance, GDP, and average payroll measures before using them as covariates in the model.

An additional complication in this data is that, for most airports, there is no direct traffic between them.    
Using ordinary least squares on only the positive observations results in an inconsistent estimator of $\boldsymbol{\beta}$ \citep{wooldridge2001econometric} (Chapter 16.3). Therefore, instead, we estimated both the regression coefficients $\boldsymbol{\beta}$ and covariance parameters using a maximum pseudo-likelihood approach \citep{besag1975statistical, arnold1991pseudolikelihood,strauss1990pseudolikelihood}.  We use techniques similar to \citet{fieuws2006pairwise} and \citet{solomon2017pseudo} for longitudinal observations on the same individual, but modified the approach to account for network structure.

A required input to the pseudo-likelihood estimation procedure is known or estimated block memberships. A preliminary estimate of $\widehat{\boldsymbol{\beta}}_E$ was obtained assuming exchangeable errors. These coefficient estimates were then used to compute residuals $r_{ij}=y_{ij}-\widehat{\boldsymbol{\beta}}_E \boldsymbol{X}_{ij}, \;\; \forall y_{ij}>0$, and Algorithm 2 was performed on the residuals for positive observations to obtain block membership estimates. Given the block memberships, the covariate effects $\boldsymbol{\beta}$ and the block-exchangeable covariance parameters could be simultaneously estimated using the pseudo-likelihood framework. 

\begin{figure}[]
\centering
  \includegraphics[width=.5\columnwidth]{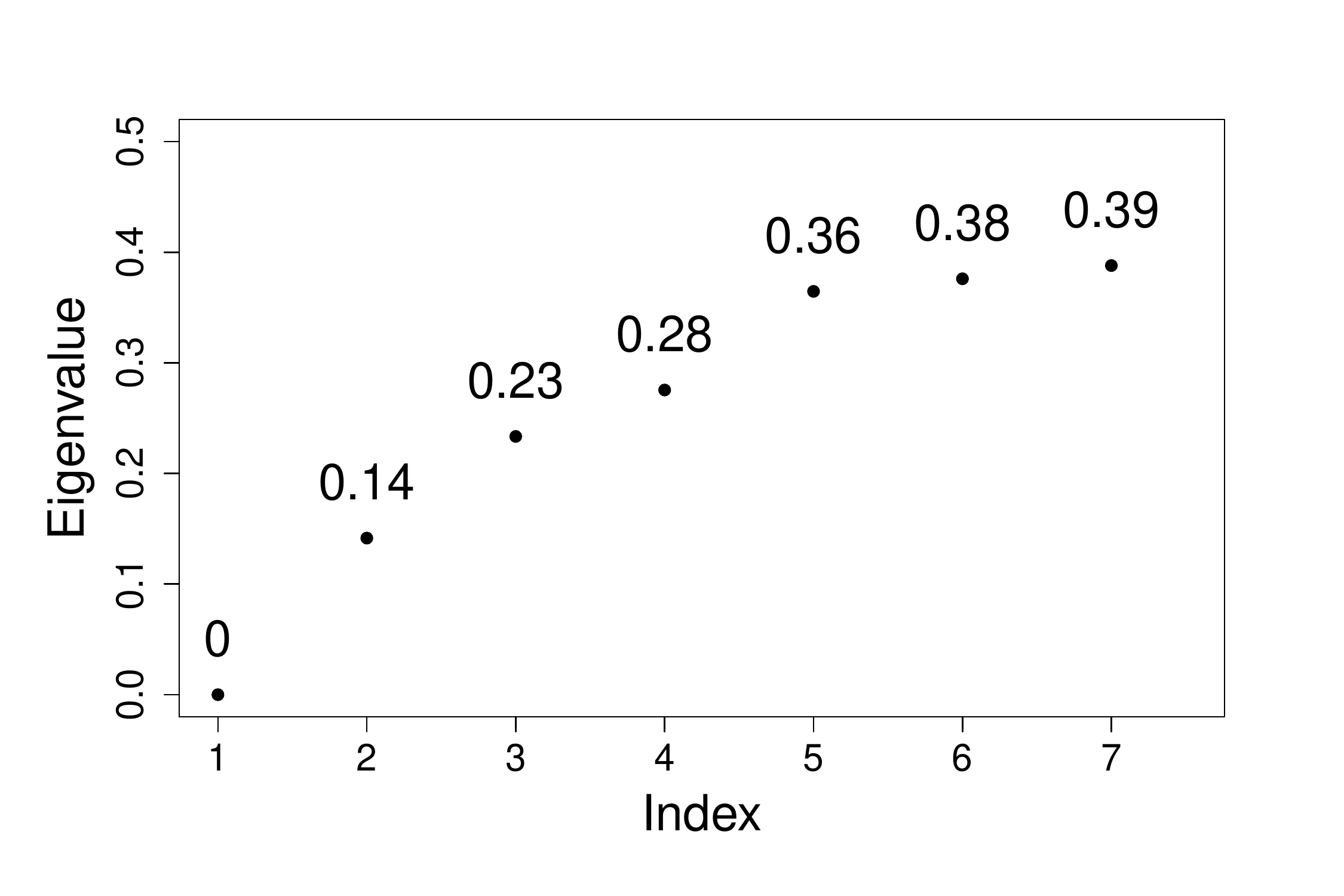}
  \caption{Smallest eigenvalues of the Laplacian matrix in increasing order}
  \label{fig:eigengap}
\end{figure}

To numerically optimize the pseudo-likelihood, we used \texttt{optim} in R, with \texttt{method="L-BFGS-B"}. We do not set bounds on $\boldsymbol{\beta}$, but did place a lower bound of $1e^{-2}$ for all variance parameters and a bound of $[-0.9,0.9]$ for all correlation parameters. We used the eigengap method, which locates a large gap between two subsequent eigenvalues, to choose the number of blocks $B$. Figure \ref{fig:eigengap} shows the smallest seven eigenvalues in increasing order. The gap between $\lambda_2$ and $\lambda_3$ is larger than the gap between $\lambda_3$ and $\lambda_4$, suggesting that $B=2$ is a reasonable choice. Figure \ref{fig:eigengap} also shows that the gap between $\lambda_4$ and $\lambda_5$ is large, suggest $B=4$ may also be appropriate.  When running the spectral clustering algorithm with $B=3$ and $B=4$, the smallest block size contained just two airports. Therefore, we proceeded with fitting a block-exchangeable covariance estimator with two blocks, which resulted in one block estimated to have 49 airports, and the other having 524 airports. Full details are provided in the supplementary materials.

Table \ref{table:airport} shows $95\%$ confidence intervals of coefficients using exchangeable estimator and block-exchangeable estimator. We see that the distance between airports is negatively associated with number of passenger seats, while GDP and average payroll of both departure and arrival airports' counties are positively associated with traffic between airports. Compared to the exchangeable estimator, the block-exchangeable estimator returns large effects of economic factors on airport traffic. 

\vspace{4mm}
\begin{table}[h]
\centering
\begin{tabular}{| c| c| c| c| }
\hline 
 & intercept & distance & GDP$_i$  \\ \hline
Exch & (-29.00, -28.94)  &  (-7.27, -7.21) & (1.89,1.95)   \\ \hline
B-E & (-26.25, -26.18) &   (-6.99, -6.92) & (3.38, 3.46)    \\  \hline
\end{tabular}
~\\~\\
\begin{tabular}{| c| c| c| c| }
\hline 
 & GDP$_j$ & payroll$_i$ & payroll$_j$ \\ \hline
Exch &  (1.89, 1.95) &  (1.42, 1.48) &   (1.41, 1.47) \\ \hline
B-E &  (3.44, 3.51) &  (2.99, 3.06) &  (2.96, 3.04)   \\  \hline
\end{tabular}
\caption{$95\%$ confidence intervals of coefficients using exchangeable estimator and block-exchangeable estimator. Distance represents the standardized great circle distance between two airports, GDP$_i$ and GDP$_j$ denote standardized GDP for departure and arrival airport, respectively, and payroll$_i$ and payroll$_j$ denote standardized average payroll for departure and arrival airport, respectively.}
\label{table:airport}
\end{table}

\section{Discussion}
In this paper, we propose a novel block-exchangeable estimator to estimate the standard errors of regression coefficients, assuming block-exchangeability. Our proposed estimator bridges the gap between the existing dyadic clustering estimator, where no distributional assumptions are made, and the exchangeable estimator, where the joint distribution of errors are assumed to be exchangeable. Through theory and simulations, we have shown that when latent block memberships are correlated with the generative process of the covariates, our block-exchangeable estimator outperforms the exchangeable estimator by having less bias of coverage, and outperforms dyadic clustering estimator by having less variance. 

Because there may not exist a link between every pair of actors in real network data, we extend our estimation algorithms to a case where we assume relational observations are zero left censored. In contrast to the method of moments approach we propose for uncensored data, we use a maximum pseudo-likelihood approach to estimate both the regression coefficients and covariance parameters simultaneously. Although maximum pseudo-likelihood estimates are less preferable to maximum likelihood estimates, using the likelihood directly is not computationally feasible in this censored data setting.

Although we focus our discussion on the impact of block dependence on inference for regression coefficients, possibly equally as interesting, is how the covariance structure, and inferred block structure, is impacted by the inclusion of covariates. In many settings--namely where a researcher is conducting experiments on graphs or wants to make causal claims--the role of covariates is often paramount. As an example, if a researcher can identify covariates that induce very strong residual block structure, these blocks may suffice for units for randomized in a causal inference study.

There are a few limitations of our work, and we discuss them here. We consider linear regression and continuous relational observations on a fully connected network, and assume actors are sampled randomly. A  future direction for this work includes extending it to respondent-driven samples.  Extending this approach to the generalized linear model framework is unfortunately nontrivial due to the coupling of the relation mean and variance in non-Gaussian link functions.
Additionally, if the block sizes are unbalanced, the variance of the estimated parameters associated with the smallest block is presumably largest.  Comparing the performances of different estimators at various levels of unbalanced block size is a direction for future study.  Finally, in the case of unknown block memberships, Algorithm 2 attempts to identify memberships based on similarities between the distribution of actor residual products. Computing these similarity scores is computationally intensive and in our examples, required a matter of hours using a standard laptop with codes written in R and not optimized for efficiency. 

\section*{Acknowledgements}

We thank the anonymous reviewers that provided feedback on our work.  This work was partially supported by NSF awards IOS-1856229 and DMS-1737673, as well as the National Institute Of Mental Health of the National Institutes of Health under Award Number DP2MH122405. 

\pagebreak
\centerline{\huge Appendix}
\counterwithin{figure}{section}
\counterwithin{table}{section}
\counterwithin{equation}{section}

\appendix

\section{Proof of Theorem 5.1}
\label{sec:proof}
We first restate Theorem 5.1, then provide a complete proof.

Assume (a) the error vector satisfies the block-exchangeability assumption, with two blocks of sizes $n_1$ and $n_2$, (b) $\boldsymbol{X}$ is a full rank $(n(n-1) \times 2)$ matrix, (c) covariates $\{X_{ij}\}$ are independent and identically distributed, (d) the fourth moment of the errors and covariates are bounded, (e) errors $\Xi$ and $\boldsymbol{X}$ are independent, and (f) the number of blocks $B$ is $\mathcal{O}(1)$.
As $n_1 \rightarrow \infty,n_2 \rightarrow \infty $, and $ n_1/n_2 \rightarrow \alpha$, where $\alpha$ is a constant such that $0 < \alpha < \infty$,
    \begin{equation}
n \left( \hat{V}_B(\hat{\boldsymbol{\beta}})- \hat{V}_E(\hat{\boldsymbol{\beta}}) \right)  \overset{p}{\to} c (\boldsymbol{X}).
 \end{equation}
 where $c(\boldsymbol{X})$ is a weighted linear combination of the differences between the true block exchangeable parameters and corresponding exchangeable parameters when the block exchangeable parameters are appropriately averaged within configuration type and convergence is pointwise. Furthermore, when $X_{ij}$ is independent of $g_i$ and $g_j$, $c(\boldsymbol{X})$ = $\boldsymbol{0}$ and thus the estimators are asymptotically equivalent.~\\
 
 We now proceed with the proof.  We begin by defining $c(\boldsymbol{X})$:
 \begin{align*} 
 c(\boldsymbol{X})&= \sum_{M,q \in Q_M} f_{M,q}  (M_q-M) \\
 &= \sum\limits_{u,v \in \{1,2\}}  f_{\sigma^2,(u,v)}   ({\sigma}_{(u,v)}^2-{\sigma}^2 ) +  \sum\limits_{u,v \in \{1,2\}} f_{\phi_A,\{u,v\}}   ({\phi}_{A,\{u,v\})}-{\phi}_A ) + ...
 \numberthis
 \label{eq:main-proof}
\end{align*} 
where $f_{M,q}$ are functions of $\boldsymbol{X}$. More specifically, given $M$ and $q$, $f_{M,q}$ is a function of elements in the set $\{ [X_{ij},X_{kl}] |  [(i,j),(k,l)] \in \Phi_{M,q} \}$. The parameter
 \begin{equation}
    \sigma^2=\frac{n_1(n_1-1) \sigma_{(1,1)}^2  +n_2(n_2-1) \sigma_{(2,2)}^2  +  n_1n_2 (\sigma_{(1,2)}^2 +\sigma_{(2,1)}^2  ) }{n(n-1)}
 \end{equation}

We now present a proof of Theorem 5.1.
\begin{align*}
   & n \left(  \hat{V}_B(\hat{\boldsymbol{\beta}}) -\hat{V}_E(\hat{\boldsymbol{\beta}}) \right)  \\
&=\left(\boldsymbol{X}^T\boldsymbol{X} \right)^{-1} \boldsymbol{X}^T (\widehat{\Omega}_B -\widehat{\Omega}_E) X(\boldsymbol{X}^T\boldsymbol{X})^{-1}    \\
   & = \frac{n}{n^2(n-1)^2} \left(\frac{\boldsymbol{X}^T\boldsymbol{X} }{ n(n-1) } \right)^{-1}
   \left(\sum\limits_{M \in \mathcal{M}} \sum\limits_{q \in Q_M} \frac{ \sum\limits_{(j,k),(m,n) \in  \Phi_{M,q} } \boldsymbol{X}_{jk} \boldsymbol{X}_{mn}^T \left( \widehat{M}_{q}-\widehat{M}  \right) |\Phi_{M,q}| }{|\Phi_{M,q}|} \right)
   \left(\frac{\boldsymbol{X}^T\boldsymbol{X} }{ n(n-1) } \right)^{-1} \\
   & = \sum\limits_{M \in \mathcal{M}} \sum\limits_{q \in Q_M} \frac{|\Phi_{M,q}|}{n(n-1)^2} \left( \widehat{M}_{q}-\widehat{M}   \right)\left(\frac{\boldsymbol{X}^T\boldsymbol{X} }{ n(n-1) } \right)^{-1} 
  \left(  \frac{\sum\limits_{(j,k),(m,n) \in  \Phi_{M,q} }\boldsymbol{X}_{jk} \boldsymbol{X}_{mn}^T}{|\Phi_{M,q}|} \right)
   \left(\frac{\boldsymbol{X}^T\boldsymbol{X} }{ n(n-1) } \right)^{-1} \\
   & = \sum\limits_{M \in \mathcal{M}} \sum\limits_{q \in Q_M} \frac{c_{M,q}\cdot |\Phi_M| }{n(n-1)^2}\left(\widehat{M}_{q}-\widehat{M}  \right) h_{M,q}(\boldsymbol{X}) \\
    & = \sum\limits_{M \in \mathcal{M}} \sum\limits_{q \in Q_M} c'_{M} c_{M,q}\left(\widehat{M}_{q}-\widehat{M}  \right) h_{M,q}(\boldsymbol{X})
   \numberthis
   \label{eq:proof_1}
\end{align*}
where $c'_{M}=\frac{|\Phi_{M}|}{n(n-1)^2} $, $c_{M,q}$ is the proportion of dyad pairs with configuration $M$ and block specification $q$ over all dyad pairs with configuration $M$, and $h_{M,q}$  contains the remaining terms which are functions of $\boldsymbol{X}$. Because we assume $B$ is $\mathcal{O}(1)$, each $|\Phi_{M}|$ is at most $\mathcal{O}(n^3)$, so each $c'_{M}  \rightarrow d_{M}$ for some constant $d_{M}$. 
\citet{marrs2017standard}  (Eq.27) show that \\
$h_{M,q}(\boldsymbol{X}) \overset{p}{\to} h'_{M,q}(\boldsymbol{X}) = \\
\begin{cases}
       \E[\boldsymbol{X}_{jk}\boldsymbol{X}_{jk}^T]^{-1}\E[\boldsymbol{X}_{jk}\boldsymbol{X}_{jk}^T  |  (j,k) \in \Phi_{\sigma^2,q}]\E[\boldsymbol{X}_{jk}\boldsymbol{X}_{jk}^T]^{-1}, & \text{for } M = \sigma^2 \\
    \E[\boldsymbol{X}_{jk}\boldsymbol{X}_{jk}^T]^{-1}\E[\boldsymbol{X}_{jk}\boldsymbol{X}_{mn}^T  |  (j,k),(m,n) \in \Phi_{M,q}]\E[\boldsymbol{X}_{jk}\boldsymbol{X}_{jk}^T]^{-1}, & \text{for } M \in \mathcal{M}\setminus \sigma^2
\end{cases} $ \\
We have shown $c_{M,q}$ and $h_{M,q}$ both converge in probability to constants. So the only part left in Equation \ref{eq:proof_1} is $ \left(\widehat{M}_{q}-\widehat{M}  \right)$.
Previous work ~\citep{marrs2017standard} has shown that 
\begin{equation}
\widehat{M}_q \overset{p}{\to} M_q \text{ and } \widehat{M} \overset{p}{\to} M,
\end{equation}
where
\begin{equation}
M=\frac{\sum\limits_{q \in Q_M} M_q \cdot|\Phi_{M,q}| }{\sum\limits_{q \in Q_M} |\Phi_{M,q}|} =\sum\limits_{q \in Q_M} M_q \cdot c_{M,q} 
\end{equation}
Thus, by Slutsky's theorem, 
\begin{equation}
n \left(  \hat{V}_B(\hat{\boldsymbol{\beta}}) -\hat{V}_E(\hat{\boldsymbol{\beta}}) \right)\overset{p}{\to} \sum\limits_{M \in \mathcal{M}} \sum\limits_{q \in Q_M}  \left({M}_{q}-{M}  \right) f_{M,q}(\boldsymbol{X}),    
\end{equation}
where $f_{M,q}=c_{M,q}\cdot d_{M} \cdot h'_{M,q}(\boldsymbol{X})$ is a constant when distribution of $\boldsymbol{X}$ is known, and $M_q$ is the true parameter in $\Omega_B$.
When the distribution of $\boldsymbol{X}$ is independent of block membership, we have $f_{M,q}(\boldsymbol{X}) = f_{M}(\boldsymbol{X}) \;\; \forall q$. In addition, $\sum\limits_{q \in Q_M} c_{M,q}=1 \; \forall M$. Therefore,
\begin{align*}
 n \left(  \hat{V}_B(\hat{\boldsymbol{\beta}}) -\hat{V}_E(\hat{\boldsymbol{\beta}}) \right) &  \overset{p}{\to}  \sum\limits_{M \in \mathcal{M}} d_M f_{M}(\boldsymbol{X})\sum\limits_{q \in Q_M} c_{M,q} \left({M}_{q}-{M}  \right) \\
 & = \sum\limits_{M \in \mathcal{M}} d_M f_{M}(\boldsymbol{X}) 
\left( \sum\limits_{q \in Q_M} c_{M,q}{M}_{q} - \sum\limits_{q \in Q_M} c_{M,q}M  \right) \\
 & = \sum\limits_{M \in \mathcal{M}} d_M f_{M}(\boldsymbol{X}) 
 \left(M-M  \right) = 0
 \numberthis
\end{align*}
Therefore, we have shown that when $\boldsymbol{X}$ is independent of $g$, $n \left(  \hat{V}_B(\hat{\boldsymbol{\beta}}) -\hat{V}_E(\hat{\boldsymbol{\beta}}) \right) \overset{p}{\to} 0$.

In the case of two blocks, 
\begin{align*}
& n \left(  \hat{V}_B(\hat{\boldsymbol{\beta}}) -\hat{V}_E(\hat{\boldsymbol{\beta}}) \right)  
 = \sum\limits_{u,v \in \{ 1,2 \}} \left( \sigma_{(u,v)}^2 -\sigma^2 \right) f_{\sigma^2, (u,v)} (\boldsymbol{X}) \\
& +  \sum\limits_{u,v \in \{ 1,2 \}} \left( \phi_{A,\{u,v\}} -\phi_A \right) f_{\phi_A, (u,v)} (\boldsymbol{X}) 
 +  \sum\limits_{u,v,w \in \{ 1,2 \}} \left( \phi_{B,(u,\{v,w\})} -\phi_B \right) f_{\phi_B, (u,\{v,w\})} (\boldsymbol{X}) \\
& +  \sum\limits_{u,v,w \in \{ 1,2 \}} \left( \phi_{C,(u,\{v,w\})} -\phi_C \right) f_{\phi_C, (u,\{v,w\})} (\boldsymbol{X}) 
 +  \sum\limits_{u,v,w \in \{ 1,2 \}} \left( \phi_{D,(uv,w)} -\phi_D \right) f_{\phi_D, (u,v,w)} (\boldsymbol{X}), 
\end{align*}
where
\begin{itemize}
    \item $\displaystyle \sigma^2=\frac{n_1(n_1-1) \sigma_{(1,1)}^2  +n_2(n_2-1) \sigma_{(2,2)}^2  +  n_1n_2 (\sigma_{(1,2)}^2 +\sigma_{(2,1)}^2  ) }{n(n-1)}$
    \item $\displaystyle \phi_A=\frac{n_1(n_1-1) \phi_{A,\{1,1\}} +n_2(n_2-1) \phi_{A,\{2,2\}}  +  2n_1n_2\phi_{A,\{1,2\}} }{n(n-1)}$
    \item $\displaystyle \phi_B=\frac{n_1(n_1-1)(n_1-2) \phi_{B(1,\{1,1\})}  +2n_1(n_1-1)n_2 \phi_{B(1,\{1,2\})}
    +n_1n_2(n_2-1) \phi_{B(1,\{2,2\})}}{n(n-1)(n-2)}$\vspace{2mm}\\ 
    +$\displaystyle \frac{
    +n_2(n_2-1)(n_2-2) \phi_{B(2,\{2,2\})}
    +2n_2n_1(n_2-1) \phi_{B(2,\{1,2\})}
    +n_2n_1(n_1-1) \phi_{B(2,\{1,1\})}
    }{n(n-1)(n-2)}$
    \item $\displaystyle \phi_C=\frac{n_1(n_1-1)(n_1-2) \phi_{C(1,\{1,1\})}  +2n_1(n_1-1)n_2 \phi_{C(1,\{1,2\})}
    +n_1n_2(n_2-1) \phi_{C(1,\{2,2\})}}{n(n-1)(n-2)}$\vspace{2mm}\\ 
    +$\displaystyle \frac{
    +n_2(n_2-1)(n_2-2) \phi_{C(2,\{2,2\})}
    +2n_2n_1(n_2-1) \phi_{C(2,\{1,2\})}
    +n_2n_1(n_1-1) \phi_{C(2,\{1,1\})}
    }{n(n-1)(n-2)}$
    \item $\displaystyle \phi_D=\frac{n_1(n_1-1)(n_1-2) \phi_{D(1,1,1)}  +n_1(n_1-1)n_2 \phi_{D(1,1,2)}+n_1(n_1-1)n_2 \phi_{D(1,2,1)}}{n(n-1)(n-2)}$ \vspace{2mm}\\ 
    +$\displaystyle \frac{ n_1n_2(n_2-1) \phi_{D(1,2,2)}
    +n_2(n_2-1)(n_2-2) \phi_{D(2,2,2)}+n_2n_1(n_2-1) \phi_{D(2,1,2)} }{n(n-1)(n-2)}$ \vspace{2mm}\\ 
    +$\displaystyle \frac{n_2n_1(n_2-1) \phi_{D(2,2,1)}
    +n_2n_1(n_1-1) \phi_{D(2,1,1)}}{n(n-1)(n-2)}$.
\end{itemize}

\section{Additional simulation details}
In this section, we provide additional details about the simulation presented in Section 6 of the manuscript.  To begin, take the generative model as: 
\vspace{-2mm}
\begin{center}
$y_{ij}=\beta_0+\beta_1X_{ij}+\xi_{ij}$,
$\xi_{ij}=a_i+b_j+z_i^Tz_j + \gamma_{(ij)}+\epsilon_{ij},$
$(a_i,b_i)|g_i \sim N_2(0,\Sigma_{ab,g_i})$;
$\Sigma_{ab,g_i} = \begin{pmatrix} 
\sigma_{a,g_i}^2 & \rho_{ab}\sigma_{a,g_i}\sigma_{b,g_i}  \\ 
 \rho_{ab}\sigma_{a,g_i}\sigma_{b,g_i} & \sigma_{b,g_i}^2
\end{pmatrix} $;  \\
%\vspace{4mm}
$z_i |g_i \sim N_d\left( 0, \sigma_{z,g_i}^2 I_d \right)$; $\epsilon_{ij}\sim N(0, \sigma_{\epsilon}^2)$; \\
%\vspace{4mm}
$\gamma_{(ij)}=\gamma_{(ji)}|g_i,g_j \sim (0, \sigma_{\gamma,\{g_i,g_j\}}^2)$.
\end{center}

Under the generative model, the variance and covariances are: \\ 
{
\begin{itemize}
\item Var$(\xi_{ij})= \sigma_{a, g_i}^2 + \sigma_{b,g_j}^2 + d \sigma_{z,g_i}^2 \sigma_{z,g_j}^2 +\sigma_{\gamma, \{g_i,g_j\}}^2+\sigma_{\epsilon;}^2 $
\item Cov$(\xi_{ij}, \xi_{ji})=\rho_{ab} \sigma_{a,g_i}\sigma_{b,g_i}  +\rho_{ab} \sigma_{a,g_j}\sigma_{b,g_j} +  d \sigma_{z,g_i}^2 \sigma_{z,g_j}^2+\sigma_{\gamma, \{g_i,g_j\};}^2$ 
\item Cov$(\xi_{ij}, \xi_{il})= \sigma_{a,g_i;}^2$ 
\item Cov$(\xi_{ij},\xi_{kj})=  \sigma_{b,g_j;}^2$
\item Cov$(\xi_{ij}, \xi_{ki}) = \rho_{ab} \sigma_{a,g_i}\sigma_{b,g_i.} $ 
\end{itemize}
}
We recognize that the error vector satisfies the block-exchangeability by making the observation that $\text{Cov}(\xi_{ij},\xi_{kl})= \text{Cov}(\xi_{\pi(i)\pi(j)},\xi_{\pi(k)\pi(l)})$ with $g_i=g_{\pi(i)}, g_j=g_{\pi(j)}, g_k=g_{\pi(k)}, \text{ and } g_l=g_{\pi(l)}$. However, this does not correspond to the most general form of the covariance matrix $\Omega_{B}$ that satisfy block-exchangeability. For example, under the error generating model, $\text{Cov}(\xi_{ij}, \xi_{il})$ takes $B$ parameters, compared to $B^2(B+1)/2$ in the most general form in Table 1 in the main document. 

\begin{figure}[ht]
\vskip 0.2in
\begin{center}
\centerline{\includegraphics[width=1\columnwidth]{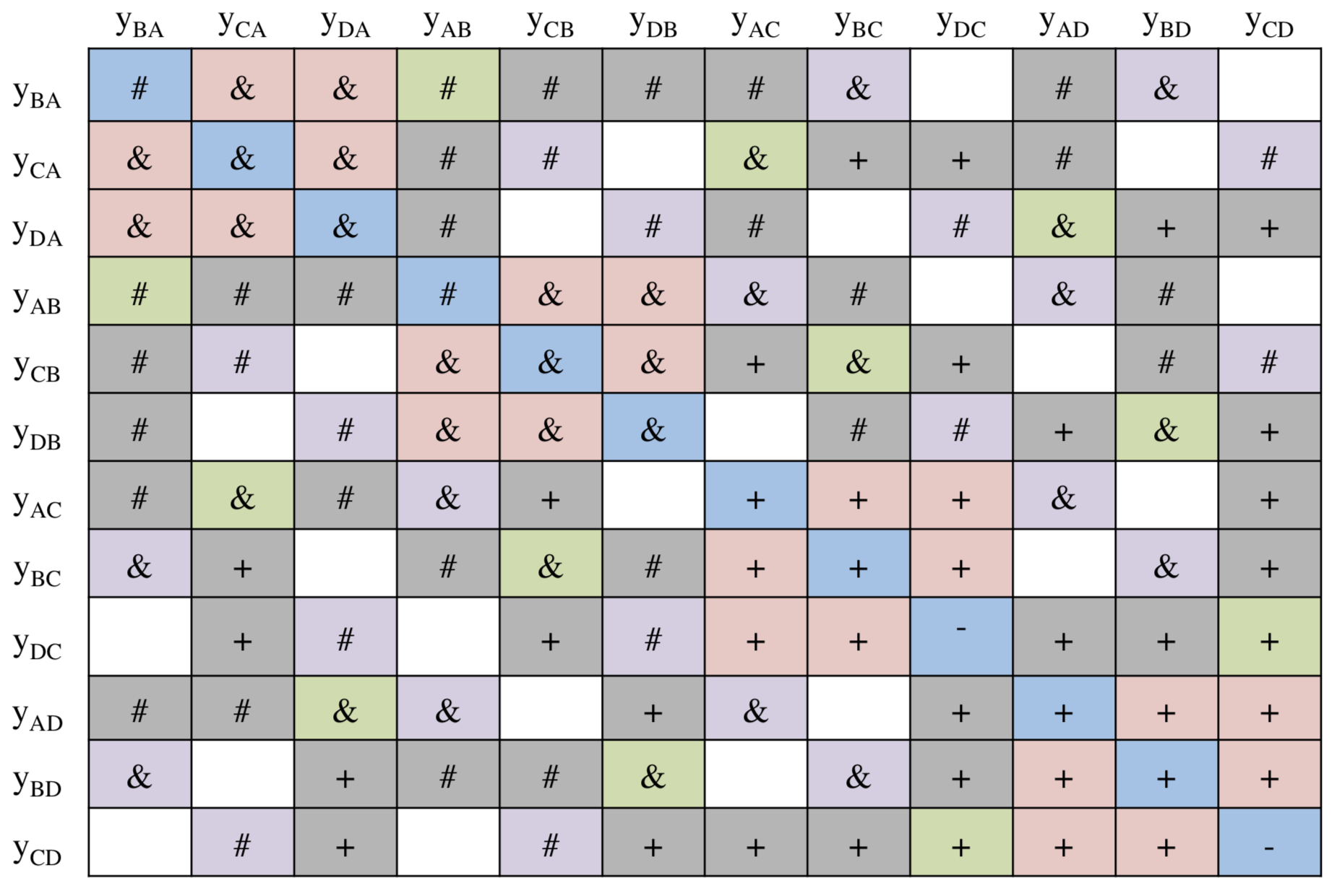}}
        \caption{Visualization of covariance matrix $\Omega$ under the error generating model used in simulation. Entries shaded with the same color and symbol share the same parameter value, and a white box indicates a covariance of zero.}
        \label{fig:generative_covariance}
\end{center}
\vskip -0.2in
\end{figure}

Figure \ref{fig:generative_covariance} shows a visualization of the covariance matrix $\Omega_B$ under the error generating model. Entries shaded with the same color and symbol share the same covariance value. Compared to Figure 1 in the main text, the error generative model does not correspond to the most general formulation of block-exchangeability covariance structure. For example, $cov(\xi_{ij},\xi_{ik})$ can take B values under the error generating model, but on the order of $B^3$ with the most general formulation.

We generate three types of covariates, each having three sub-cases regarding the correlation between the covariate and block membership:
\begin{enumerate}
    \item $X_{ij,1}=\mathbbm{1}_{X_i=X_j} $, where $X_i \sim \text{Bernoulli}(p_{g_i}) $ and 
    \begin{enumerate}
        \item $p_{g_i}$ is uncorrelated with $g_i$, i.e., $p_{g_i}$ is a fixed number
        \item $p_{g_i}|g_i=2>p_{g_j}|g_j=1>0.5 $, which suggests that high $\text{Var}(X_{ij,1})$ is associated with high $\text{Var}(\xi_{ij})$
        \item $p_{g_i}|g_i=1>p_{g_j}|g_j=2>0.5 $ , which suggests that high $\text{Var}(X_{ij,1})$ is associated with low $\text{Var}(\xi_{ij})$
    \end{enumerate}
    \item $X_{ij,2}=|X_i-X_j| $, where $X_i \sim \text{N}(0,\sigma_{g_i}) $ and 
    \begin{enumerate}
        \item $\sigma_{g_i}$ is uncorrelated with $g_i$, i.e., $\sigma_{g_i}$ is a fixed number
        \item $\sigma_{g_i}|g_i=1 > \sigma_{g_i}|g_i=2$, which suggests that high $\text{Var}(X_{ij,2})$ is associated with high $\text{Var}(\xi_{ij})$.
        \item $\sigma_{g_i}|g_i=1 < \sigma_{g_i}|g_i=2$, which suggests that high $\text{Var}(X_{ij,2})$ is associated with low $\text{Var}(\xi_{ij})$.
    \end{enumerate}
    \item $X_{ij,3} \sim N(0,\sigma_{g_i,g_j}^2)$ and 
    \begin{enumerate}
        \item $\sigma_{g_i,g_j}$ is uncorrelated with $g_i, g_j$, i.e., $\sigma_{g_i,g_j}$ is a fixed number
        \item $\sigma_{g_i,g_j}|g_i=1,g_j=1 > \sigma_{g_i,g_j}|g_i=2,g_j=2 $, which suggests that high $\text{Var}(X_{ij,3})$ is associated with high $\text{Var}(\xi_{ij,3})$
        \item $\sigma_{g_i,g_j}|g_i=1,g_j=1 < \sigma_{g_i,g_j}|g_i=2,g_j=2 $, which suggests that high $\text{Var}(X_{ij,3})$ is associated with low $\text{Var}(\xi_{ij,3})$.
    \end{enumerate}
\end{enumerate}

We set the parameters for generating covariates such that the noise to signal ratio, which is defined as the ratio of sum of squared errors over total sum of squares, is consistent across all three scenarios. Let $NTS$ denote the noise-to-signal ratio, then
\begin{center}

    $NTS_{ij}=\frac{\text{Var}(\xi_{ij})}{\text{Var}(Y_{ij})}$, where 
    $\text{Var}(\xi_{ij})=\sigma_{(g_i,g_j)}^2$ and \vspace{4mm}\\
    $\text{Var}(Y_{ij})=E(\text{Var}(Y_{ij}|X_{ij}))+\text{Var}(E(Y_{ij}|X_{ij})) = \sigma_{(g_i,g_j)}^2+ \beta_1^2 \text{Var}(X_{ij}).$
\end{center}
Therefore, for all three types of covariates:
\begin{enumerate}
    \item $X_{ij,1}=\mathbbm{1}_{X_i=X_j} $, where $X_i \sim \text{Bernoulli}(p_{g_i}) $. \vspace{4mm} \\
    $NTS_{ij}\mid g_i,g_j= \frac{\sigma_{g_i,g_j}^2}{\sigma_{g_i,g_j}^2+ \beta_1^2 p_{ij}(1-p_{ij})}$, where $p_{ij}=p_ip_j+(1-p_i)(1-p_j) $
    \item $X_{ij,2}=|X_i-X_j| $, where $X_i \sim \text{N}(0,a_{g_i}^2) $. \vspace{4mm} \\
    $NTS_{ij}\mid g_i,g_j= \frac{\sigma_{g_i,g_j}^2}{\sigma_{g_i,g_j}^2+ \beta_1^2 (a_{g_i}^2+a_{g_j}^2)(1-2/\pi) }$
    \item $X_{ij,3} \sim N(0,a_{g_i,g_j}^2)$. \vspace{4mm}\\
    $NTS_{ij}\mid g_i,g_j= \frac{\sigma_{g_i,g_j}^2}{\sigma_{g_i,g_j}^2+ \beta_1^2 a_{g_i,g_j}^2 }$
\end{enumerate}
With two blocks and equal block size, we set the equations $(\sum_{(u,v) \in \{ (1,1),(1,2),(2,1),(2,2) \}} NTS_{ij}\mid g_i=u,g_j=v )/4=0.45$ and solve for the parameters.

\section{Additional simulations: Evaluating Block Membership Estimation}
 \label{sec:block_estimation}
This section aims to show how well we recover block labels (Step 2-4 in Algorithm) as well as graphical proof of concept for why we construct the similarity metric between a pair of nodes as in Step 2 of the Algorithm. We consider a simple linear regression model with two blocks:
\begin{center}
$y_{ij}=\beta_0+\beta_1 X_{ij}+\xi_{ij}$,
\end{center}
where $X_{ij} \stackrel{i.i.d}{\sim} \mathcal{N}(0,1)$ and $g_i \in \{  1,2\}$. We vary the strength of block structure in errors and show how the algorithm recovers block membership. 

 Based on the error generating model in Section 6 of the main text, we set parameters as follows:
 \iffalse
 \begin{itemize}
\item $[\sigma_{a,1} \;\; \sigma_{a,2}]= [\sqrt{2}\alpha_1 \; \; \sqrt{2} \alpha_2]$
\item $[\sigma_{b,1} \;\; \sigma_{b,2}]=[\alpha_1 \; \;  \alpha_2]$
\item $[\sigma_{z,1} \;\; \sigma_{z,2}]=[\alpha_1 \; \;  \alpha_2]$
\item $[ \sigma_{\gamma,\{1,1\}}  \;\;   \sigma_{\gamma,\{1,2\}}  \;\;   \sigma_{\gamma,\{2,2\}} ]= [\alpha_1 \; \; \sqrt{\alpha_1 \alpha_2} \; \;  \alpha_2]$
\item $\sigma_{\epsilon}= \alpha_1$, $\rho=0.5$, and $d=2$.
\end{itemize}

Let $\alpha_2 = \alpha_1 * r$, where quantify the strength of block structure in errors. Then in terms of $r$ and $\sigma_1$, the parameters are:
\fi
 \begin{itemize}
\item $[\sigma_{a,1} \;\; \sigma_{a,2}]= [\sqrt{2}\alpha_1 \; \; \sqrt{2} r\alpha_1]$
\item $[\sigma_{b,1} \;\; \sigma_{b,2}]=[\alpha_1 \; \;  r\alpha_1]$
\item $[\sigma_{z,1} \;\; \sigma_{z,2}]=[\alpha_1 \; \;  r\alpha_1]$
\item $[ \sigma_{\gamma,\{1,1\}}  \;\;   \sigma_{\gamma,\{1,2\}}  \;\;   \sigma_{\gamma,\{2,2\}} ]= [\alpha_1 \; \; \sqrt{r} \alpha_1  \; \;  r\alpha_1]$
\item $\sigma_{\epsilon}= \alpha_1$, $\rho=0.5$, and $d=2$.
\end{itemize}
We immediately see that $r$ quantifies the strength of block structure in errors. A trivial $r=1$ suggests that there is no block structure, while an $r$ value far away from one suggests a strong block structure. As functions of $r$ and $\alpha_1$, the variance and covariances are: 

\begin{align*}
\text{Var}(\xi_{ij}) = 
  \begin{cases}
   5\alpha_1^2+2\alpha_1^4 & \text{if } g_i=1, g_j=1\\
  (r^2+r+3)\alpha_1^2+2r^2\alpha_1^4  & \text{if }  g_i=1, g_j=2 \\
    (2r^2+r+2)\alpha_1^2+2r^2\alpha_1^4  & \text{if } g_i=2, g_j=1 \\
  (4r^2+1)\alpha_1^2+2r^4\alpha_1^4   & \text{if } g_i=2, g_j=2
  \end{cases}
\end{align*}

\begin{align*}
\text{Cov}(\xi_{ij}, \xi_{ji}) = 
  \begin{cases}
   (\sqrt{2}+1)\alpha_1^2+2\alpha_1^4 & \text{if } g_i=1, g_j=1\\
   (1/\sqrt{2}+r+1/\sqrt{2}r^2)\alpha_1^2+2r^2\alpha_1^4  & \text{if }  g_i=1, g_j=2 \\
    (1/\sqrt{2}+r+1/\sqrt{2}r^2)\alpha_1^2+2r^2\alpha_1^4 & \text{if } g_i=2, g_j=1 \\
  (\sqrt{2}+1)r^2\alpha_1^2+2r^4\alpha_1^4   & \text{if } g_i=2, g_j=2
  \end{cases}
\end{align*}

\begin{align*}
\text{Cov}(\xi_{ij}, \xi_{il}) = 
  \begin{cases}
   2\alpha_1^2 & \text{if } g_i=1\\
      2r^2\alpha_1^2    & \text{if } g_i=2
  \end{cases}
\end{align*}

\begin{align*}
\text{Cov}(\xi_{ij}, \xi_{kj}) = 
  \begin{cases}
   \alpha_1^2 & \text{if } g_j=1\\
      r^2\alpha_1^2    & \text{if } g_j=2
  \end{cases}
\end{align*}

\begin{align*}
\text{Cov}(\xi_{ij}, \xi_{ki}) = 
  \begin{cases}
   1/\sqrt{2}\alpha_1^2 & \text{if } g_i=1\\
       1/\sqrt{2}r^2\alpha_1^2    & \text{if } g_i=2
  \end{cases}
\end{align*}

We perform simulation study on three values of $r$: $r=1/4, r=1/2,\text{ and } r=3/4$. Again we see that $r=1/4$ has the strongest block structure in errors, as the differences in variance and covariances between different blocks are largest. For example, $\text{Cov}(\xi_{ij}, \xi_{il} | g_i=1) - \text{Cov}(\xi_{ij}, \xi_{il} | g_i=2) = 2(1-r^2)\alpha_1^2$, and $(1-r^2)$ is a decreasing function in $r \in (0,1]$. Because all three values of $r$ are between 0 and 1, We also observe that:
\begin{itemize}
    \item $\text{Var}(\xi_{ij})|g_i=1,g_j=1>\text{Var}(\xi_{ij})|g_i=1,g_j=2>var(\xi_{ij})|g_i=2,g_j=1>\text{Var}(\xi_{ij})|g_i=2,g_j=2$
    \item $\text{Cov}(\xi_{ij}, \xi_{ji})|g_i=1,g_j=1 >\text{Cov}(\xi_{ij}, \xi_{ji})|g_i=1,g_j=2 = \text{Cov}(\xi_{ij}, \xi_{ji})|g_i=2,g_j=1 >\text{Cov}(\xi_{ij}, \xi_{ji})|g_i=2,g_j=2  $
    \item $\text{Cov}(\xi_{ij}, \xi_{il})|g_i=1 > \text{Cov}(\xi_{ij}, \xi_{il})|g_i=2$
    \item $\text{Cov}(\xi_{ij}, \xi_{kj})|g_j=1 >\text{Cov}(\xi_{ij}, \xi_{kj})|g_j=2$
    \item $\text{Cov}(\xi_{ij}, \xi_{ki})|g_i=1 > \text{Cov}(\xi_{ij}, \xi_{ki})|g_i=2$.
\end{itemize}

\subsection{Simulation Results}

\iffalse
We will show:

\begin{itemize}
    \item colored density plot of $r_{ij}r_{kl}$ for $g_i=1$ and $g_i=2$.
    \item Distribution of KS statistic/similarity between distribution of $r_{ij}r_{kl}$ for cases when $g_i=g_j$ and $g_i \neq g_j$ for each dyad 
    \item boxplot of number of misclustered nodes/n for $n=\{20,40,80\}$ using mean of 5 KS (later will do n=160)
\end{itemize}
 \fi
In this section, we provide simulation evidence for Step 2 and 3 in Algorithm 2, as well as how well we recover the block membership. Step 2 calculates the set of residual products for a specific actor and dyad configuration, and step 3 calculates the Kolmogorov-Smirnov statistic of the residual products between a pair of actors. Using simulated data, we show that the distributions of residual products for actors $i$ and $i'$ $(g_i \neq g_{i'})$ are more similar as block strength decreases, which is evidence why using the KS statistic between them is a reasonable way to construct a similarity matrix. 

\begin{figure}[H]
\centering
\includegraphics[width=1.1\linewidth]{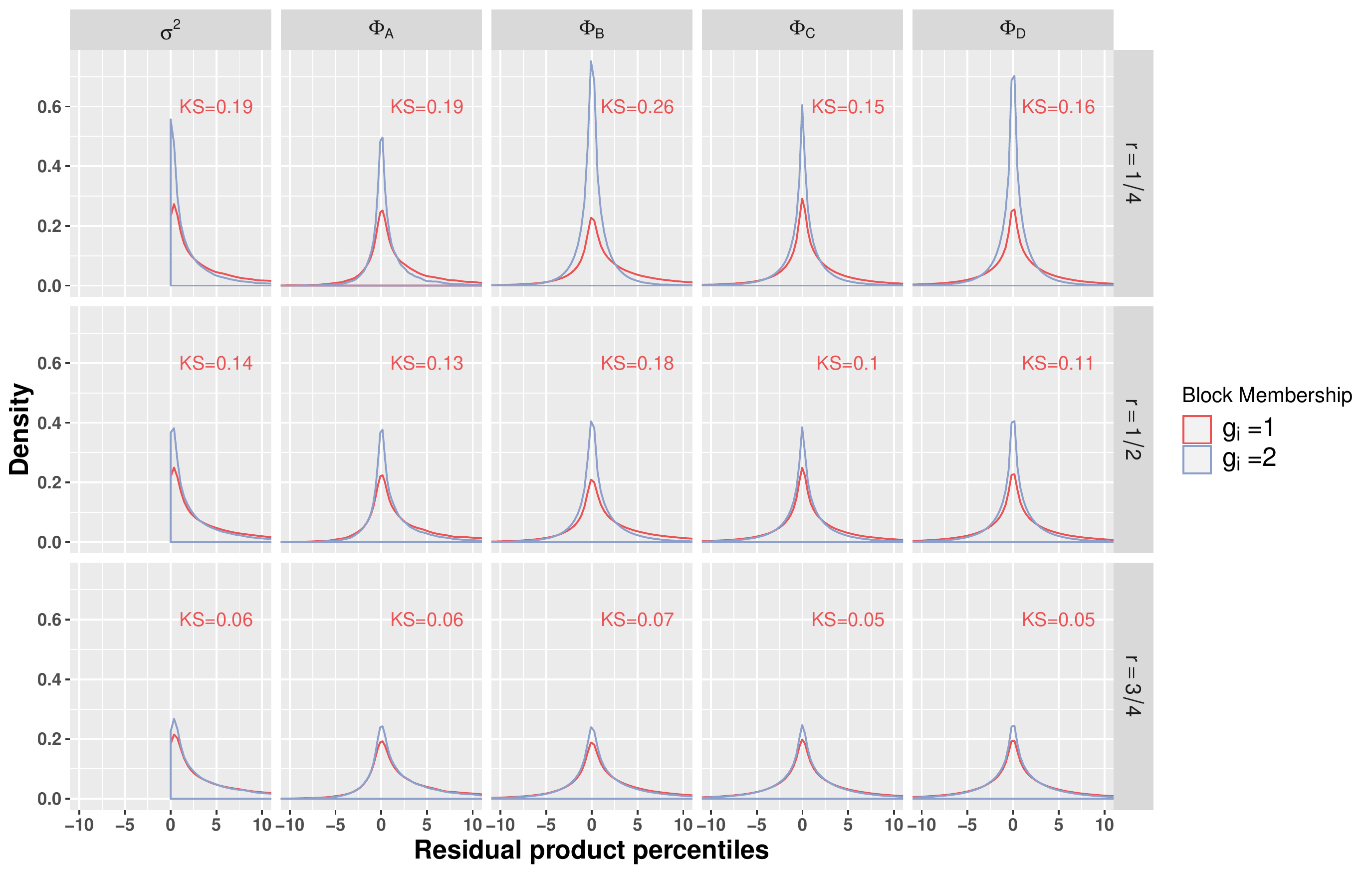}
\caption{Residual products of five dyads from 10 simulation of $n=80$. Each column represents one of the five cases $M \in \{\sigma^2, \phi_A, \phi_B,\phi_C,\phi_D\}$, and each row represents a given $r$ value. The red and blue curves represent the distribution in Block 1 and Block 2, respectively. The KS statistic on each plot is calculated between the distribution of residual products.}
\label{fig:hist_res_prod}
\end{figure}

Figure \ref{fig:hist_res_prod} shows the distribution of residual products calculated in Algorithm 2 Step 2 on each of the five cases at different values of $r$. Each column represents one of the five cases $M \in \{\sigma^2, \phi_A, \phi_B,\phi_C,\phi_D\}$, and each row represents a given $r$ value. The red and blue curves represent the distribution in Block 1 and Block 2, respectively. The densities are constructed on all actors from 10 simulations of a network of size 80. The KS statistic on each plot is calculated between the distribution of residual products. At $r=1/4$, all five plots show that the red curve is more spread out. This is because we set the simulation parameters such that variance and covariances involving actors in Block 1 is always larger than those involving Block 2. Since residual products are estimators of variance and covariances, we observe that $\forall M \in \{\sigma^2, \phi_A, \phi_B,\phi_C,\phi_D\}$, the distribution of $\boldsymbol{R}_{M,i}|g_i=1$ is more spread out. As $r$ decreases, the strength of block in errors decreases, so we observe a smaller difference between the two densities on all five cases. At $r=3/4$, the two densities coincide on $M \in \{ \phi_C,\phi_D \}$. This shows that as we have stronger block structure in errors, we have a larger difference between the distribution of residual products.

\begin{figure}[H]
\centering
\includegraphics[width=1.1\linewidth]{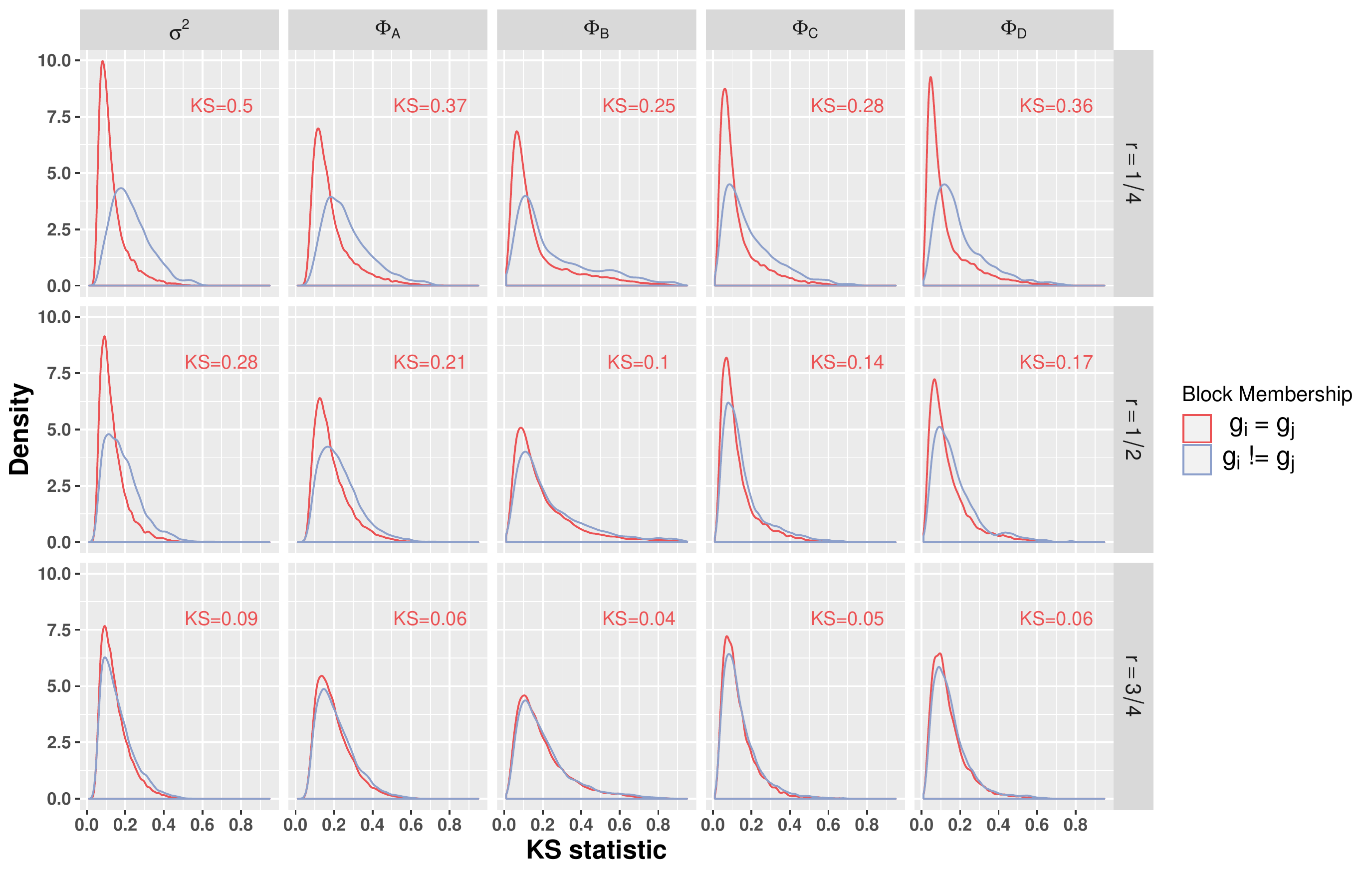}
\caption{Distribution of KS statistic between residual products of five dyads from 10 simulation of $n=80$. Each column represents one of the five cases $M \in \{\sigma^2, \phi_A, \phi_B,\phi_C,\phi_D\}$, and each row represents a given $r$ value. The red curves represent the distribution where the two actors share the same block membership ($g_i=g_j$), while the blue curves represent the distribution where the two actors are in different blocks ($g_i \neq g_j$). The KS statistic on each plot is calculated between the distribution of KS statistics.} %
\label{fig:hist_ks_stat}
\end{figure}

Figure \ref{fig:hist_ks_stat} shows the distribution of KS statistic $KS_{i,j,M}$ calculated in Algorithm 2 Step 3 on each of the five cases at different values of $r$. Each column represents one of the five cases $M \in \{\sigma^2, \phi_A, \phi_B,\phi_C,\phi_D\}$, and each row represents a given $r$ value. The red curves represent the distribution where the two actors share the same block membership ($g_i=g_j$), while the blue curves represent the distribution where the two actors are in different blocks ($g_i \neq g_j$). The densities are constructed on all actors from 10 simulations of a network of size 80. The KS statistic on each plot is calculated between the distribution of KS statistics. At $r=1/4$, we observe that the blue curve is more spread out. This is expected because the difference in distributions of residual products involving actors $i$ and that involving actor $j$ is larger when $g_i \neq g_j$, which leads to larger KS statistic between the two distributions. We also observe that when $M=\sigma^2$, the KS statistic between two distributions of KS statistic is largest, which is evidence that the distribution of $\boldsymbol{R}_{\sigma^2,i}$ is most effective in identifying whether two actors belong to the same block. At $r=3/4$, we observe that the two curves are similar. Since the block structure is not strong in errors, the distribution of $\boldsymbol{R}_{M,i}$ and $\boldsymbol{R}_{M,j}$ are not too different even when $g_i \neq g_j$.
 
\begin{figure}[H]
\centering
  \includegraphics[width=.7\linewidth]{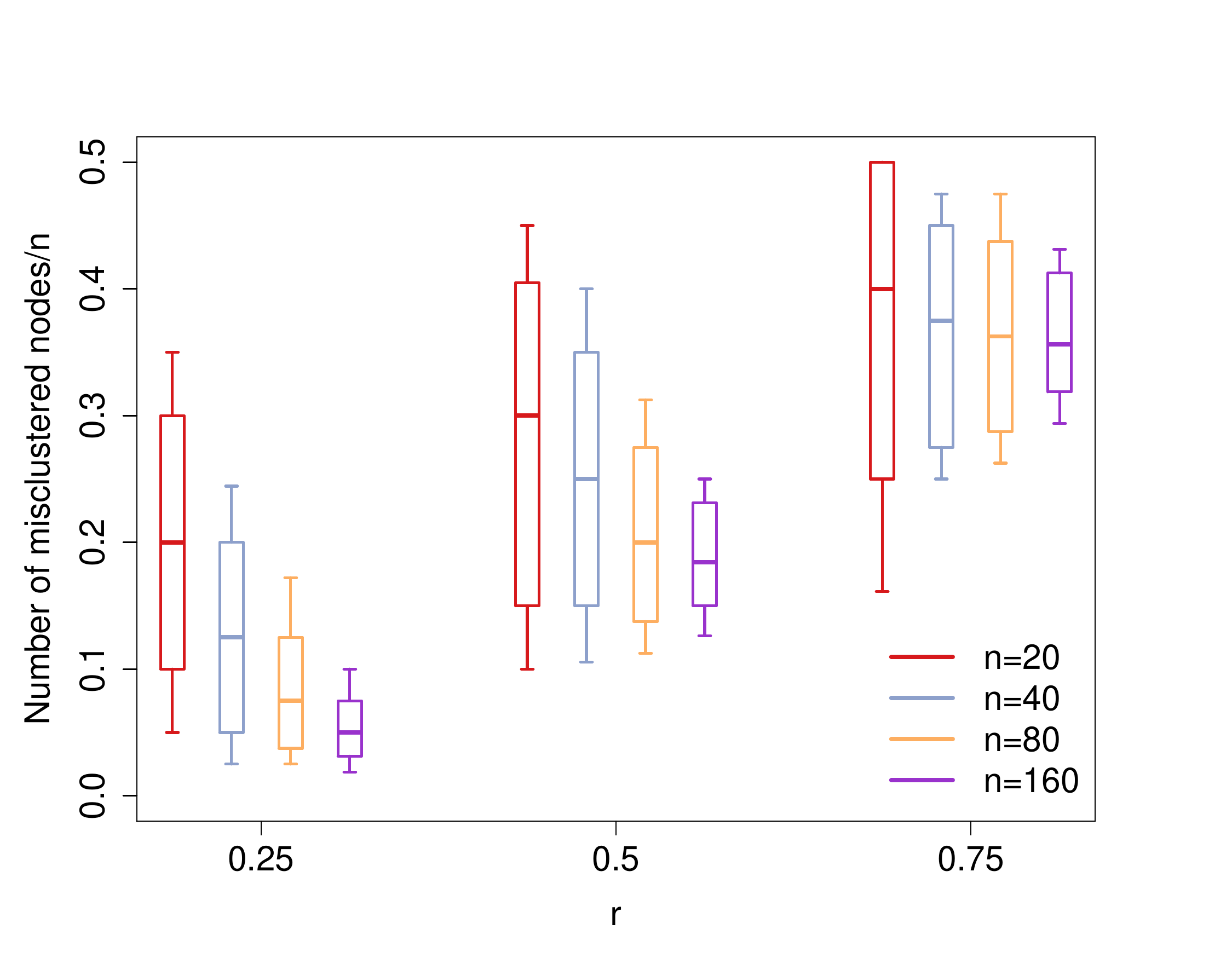}
  \caption{Number of misclustered nodes over $n$ at different $r$.    }
  \label{fig:boxplot_misclustered_nodes_vs_r}
\end{figure}

Figure \ref{fig:boxplot_misclustered_nodes_vs_r} shows the number of misclustered nodes at different values of $r$. The number of misclustered nodes is defined as $\min(\Pi_{g_i} \sum_{i=1}^n |g_i - \hat{g}_i|)$, which is the minimum number of nodes in the wrong block under permutation of the block labels. In the network of size $n$, the number of misclustered nodes ranges from 0 to $n/2$. The boxplots in Figure \ref{fig:boxplot_misclustered_nodes_vs_r} shows the distribution of the proportion of misclustered nodes, which is defined as the number of misclustered nodes over $n$, where the red, blue, yellow color represent network size $n=20, 40, 80, 160$ respectively. The line in the box is the median proportion, the boundaries of the box is 10 and 90 percentile, and the whiskers are 2.5 and 97.5 percentile. We observe that the proportion decreases with increasing $n$ and increases with increasing $r$, which shows that we recover block membership well at large network size and strong block structure in errors.

\section{Additional details on the air traffic data}
\begin{figure}[h!]
\centering
  \includegraphics[width=.45\linewidth]{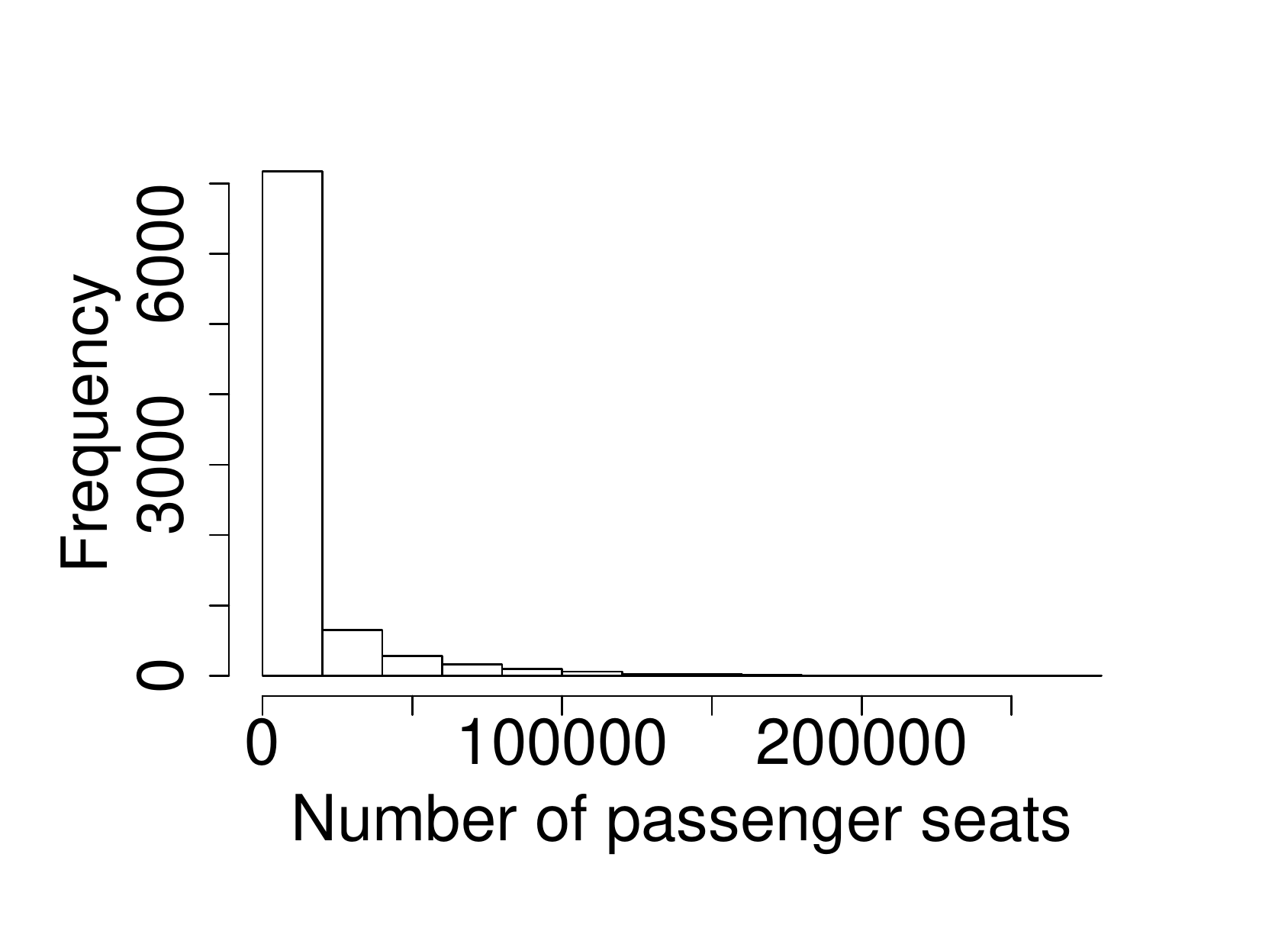}
  \hspace{.1in}
  \includegraphics[width=.45\linewidth]{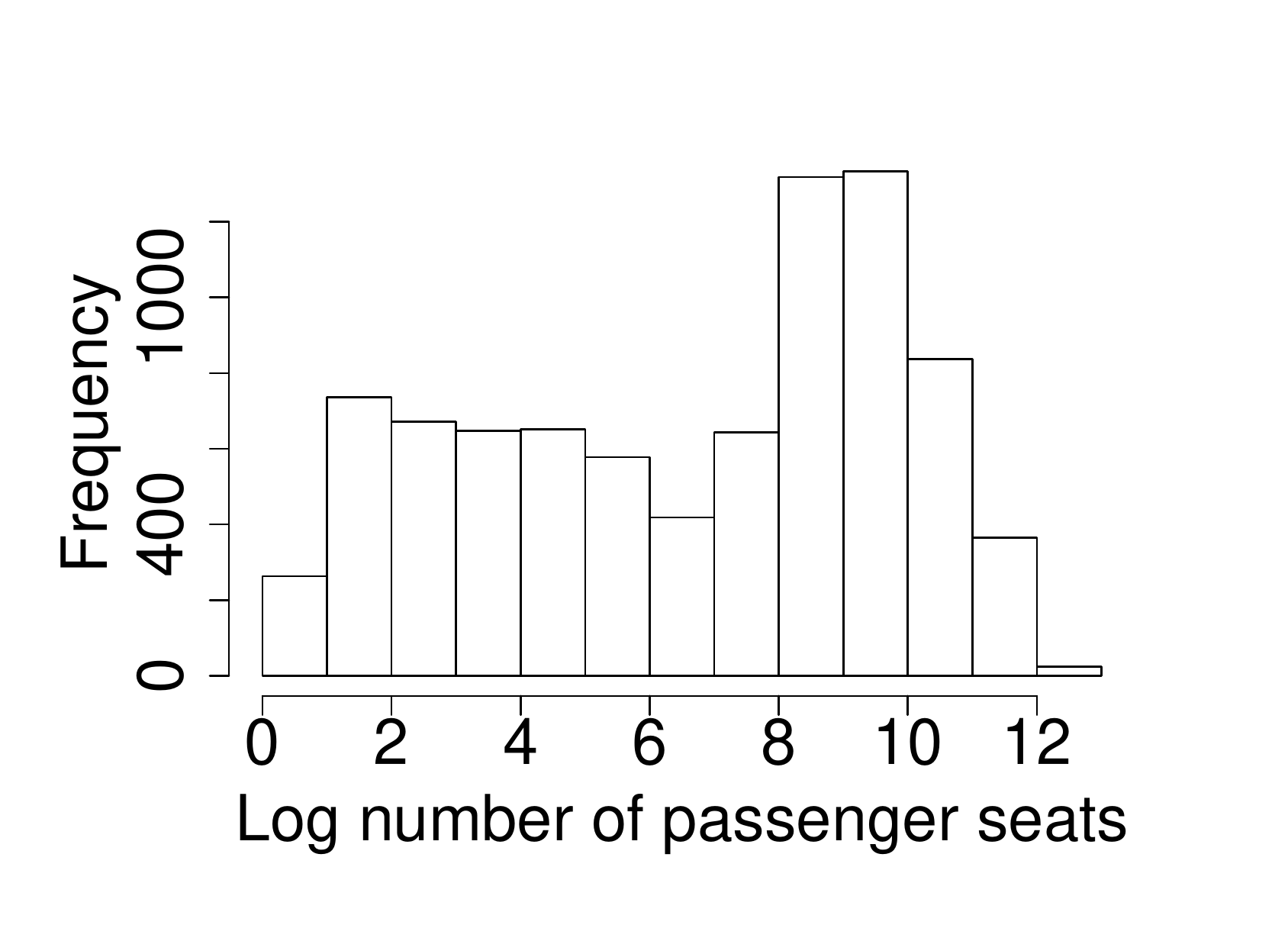}
\caption{Distribution of number of passenger seats from departure airport to arrival airport.  The left figure shows distribution of the raw number of seats.  The right figure shows the log of the number of seats when the number exceeds zero.}
\label{fig:dist_airport_passenger}
\end{figure}

In this section, we provide more details about fitting the proposed model in the context of our illustrative data example.  A challenge posed by these data is the large number of zeros that arise when there are no passenger seats from one airport to another.  Figure~\ref{fig:dist_airport_passenger} shows the distribution of passenger seats and the log number of seats between a destination for cases where the number is greater than zero.  We develop a pseudolikelihood approach to address the structure of these data. \citet{besag1975statistical} introduces pseudo-likelihood methodology using an objective function that maximizes a product of conditional densities instead of the joint likelihood, and \citet{arnold1991pseudolikelihood} shows that when using pseudo-likelihood as the objective, the parameter estimates are asymptotically normal with mean as the true parameter and covariance matrix as the sandwich estimator. In the field of network analysis, pseudo-likelihood approach has been widely used to make inference for exponential family random graph models (ERGM) (\citet{strauss1990pseudolikelihood}), due to the fact that computation of conditional densities are easier. Assume we have $n$ independent, identically distributed observed vectors $Y^{(i)}$, researchers have also used pseudo-likelihood approach as maximizing a sum of pairwise marginal log likelihoods:
\begin{equation}
 l(\boldsymbol{\theta}; Y^{(1)}, ... ,Y^{(n)} ) = \sum\limits_{i}  l(\boldsymbol{\theta}; Y^{(i)} ) \text{ ,where } \;\; l(\boldsymbol{\theta}; Y^{(i)} ) =  \sum_{s>t} \log L(Y_s^{(i)},Y_t^{(i)}; \boldsymbol{\theta}),   
\end{equation}
where $L(Y_s^{(i)},Y_t^{(i)}; \boldsymbol{\theta})$ is the likelihood of observing a pair of values $Y_s^{(i)}$ and $Y_t^{(i)}$ given parameter $\boldsymbol{\theta}$.
\citet{cox2004note} presents conditions for obtaining consistent estimates when using such approach. \citet{fieuws2006pairwise} apply this method to the case of longitudinal observations, where individual random effects lead to non-zero covariance between multiple observations on the same individual. \citet{solomon2017pseudo} extend this application to a case where observations are left-censored. Other applications of pairwise likelihood approach include \citet{kuk2000pairwise} and \citet{renard2004pairwise}.

We first present the likelihood for a pair of observations $(y_{ij},y_{kl})$ when one or both observations mat be censored. We consider a setting of relational observations left-censored at zero for the regression model below: 
\[
  y_{ij} = 
  \begin{cases}
    y_{ij}^*, & \text{if } y_{ij}^*\geq 0 \\
    0, & \text{if } y_{ij}^* < 0 
  \end{cases}
\]
where $y_{ij}^*=\boldsymbol{\beta}^T \boldsymbol{X}_{ij} + \xi_{ij}, \hspace{.1in}  i,j \in \{ 1,...,n\}, i \neq j.$
Let $\boldsymbol{\theta}$ denote the parameter vector containing $\boldsymbol{\beta}$ and covariance terms. Let $\rho_{(\epsilon_{ij},\epsilon_{kl})} =\text{Cov}(\epsilon_{ij},\epsilon_{kl}) / \sqrt{ \text{Var}(\epsilon_{ij}) \text{Var}(\epsilon_{kl})  } $ denote the correlation coefficient between $\epsilon_{ij}$ and $\epsilon_{kl}$. The likelihood of a pair of relational observations $L(y_{ij},y_{kl};\boldsymbol{\theta})$ takes one of the four following values.
\begin{itemize}
\item If $y_{ij} >0$ and $y_{kl}>0 $, then $L(y_{ij},y_{kl};\boldsymbol{\theta} )= f_{Y_{ij},Y_{kl}}(y_{ij},y_{kl}),$ \vspace{4mm}  \\ where $  \begin{pmatrix}Y_{ij} \\ Y_{kl}  \end{pmatrix} \sim \mathcal{N} \left( \begin{bmatrix} \boldsymbol{\beta}^T\boldsymbol{X}_{ij} \\ \boldsymbol{\beta}^T\boldsymbol{X}_{kl}  \end{bmatrix} ,     \begin{bmatrix}    \text{Var}(\epsilon_{ij}) &    \text{Cov}(\epsilon_{ij},\epsilon_{kl})   \\   \text{Cov}(\epsilon_{ij},\epsilon_{kl}) &  \text{Var}(\epsilon_{kl})   \end{bmatrix} \right)$.

\item If $y_{ij} >0$ and $y_{kl}=0 $, then $L(y_{ij},y_{kl};\boldsymbol{\theta} )=  f_{Y_{ij}}( y_{ij})\cdot  F_{Y_{kl} \mid Y_{ij}}(0),$  \vspace{2mm} \\ where $Y_{ij} \sim \mathcal{N} ( \boldsymbol{\beta}^T\boldsymbol{X}_{ij} ,   \text{Var}(\epsilon_{ij}) ) $\\  and
$Y_{kl} \mid Y_{ij} \sim \mathcal{N}(   \boldsymbol{\beta}^T\boldsymbol{X}_{kl}+\text{Cov}(\epsilon_{ij},\epsilon_{kl})/\text{Var}(\epsilon_{ij}) \cdot( y_{ij}- \boldsymbol{\beta}^T\boldsymbol{X}_{ij})  ,  (1-\rho_{(\epsilon_{ij},\epsilon_{kl})}^2) \cdot  \text{Var}(\epsilon_{kl})  )$

\item if $y_{ij} =0$ and $y_{kl}>0 $, then
$L(y_{ij},y_{kl};\boldsymbol{\theta} )=  f_{Y_{kl}}( y_{kl})\cdot  F_{Y_{ij} \mid Y_{kl}}(0),$  \vspace{2mm} \\ where $Y_{kl} \sim \mathcal{N} ( \boldsymbol{\beta}^T\boldsymbol{X}_{kl} ,   \text{Var}(\epsilon_{kl}) ) $\\  and
$Y_{ij} \mid Y_{kl} \sim \mathcal{N}(   \boldsymbol{\beta}^T\boldsymbol{X}_{ij}+\text{Cov}(\epsilon_{ij},\epsilon_{kl})/\text{Var}(\epsilon_{kl}) \cdot( y_{kl}- \boldsymbol{\beta}^T\boldsymbol{X}_{kl})  ,  (1-\rho_{(\epsilon_{ij},\epsilon_{kl})}^2) \cdot  \text{Var}(\epsilon_{ij})  )$

\item if $y_{ij} =0$ and $y_{kl}=0 $, then $L(y_{ij},y_{kl};\boldsymbol{\theta} )=  F_{Y_{ij}, Y_{kl}}(0,0)$  \vspace{4mm}\\
where $  \begin{pmatrix}Y_{ij} \\ Y_{kl}  \end{pmatrix} \sim \mathcal{N} \left( \begin{bmatrix} \boldsymbol{\beta}^T\boldsymbol{X}_{ij} \\ \boldsymbol{\beta}^T\boldsymbol{X}_{kl}  \end{bmatrix} ,     \begin{bmatrix}    \text{Var}(\epsilon_{ij}) &    \text{Cov}(\epsilon_{ij},\epsilon_{kl})   \\   \text{Cov}(\epsilon_{ij},\epsilon_{kl}) &  \text{Var}(\epsilon_{kl})   \end{bmatrix} \right)$.
\end{itemize}

The likelihood we present above applies to one pair of observations. To calculate the pseudo log likelihood of all pairs of observations, we have
\begin{equation}
l(\boldsymbol{\theta};Y)= \sum\limits_{i,j,k,l \in [n], i \neq j, k \neq l} \log L(y_{ij},y_{kl};\boldsymbol{\theta}). 
\label{eqn:pseudo_all}
\end{equation}
In Equation \ref{eqn:pseudo_all}, $\boldsymbol{\theta}$ contains $\boldsymbol{\beta}$ and all variance and covariance terms. With $B$ blocks and $p-1$ covariates, the total number of parameters is on the order of $p$ or $B^3$, depending on which one is larger. Estimating all parameters at the same time is too difficult for state of art optimization algorithms. The covariance between $y_{ij}$ and $y_{kl}$ depends on the dyad configuration of $[(i,j),(k,l)]$ and their block memberships. Therefore, in order to decrease the number of parameters in each numerical optimization, we decompose the likelihood into a sum of sub-likelihoods involving pairs of observations that share the same covariance. The parameter vector in each sub-likelihood contains $\boldsymbol{\beta}$, $\text{Var}(\epsilon_{ij})$, $\text{Var}(\epsilon_{ij})$, and $\text{Cov}(\epsilon_{ij},\epsilon_{kl})$. The number of parameters in each sub-likelihood is on the order $p$, which greatly reduces the difficulties for numerical optimization.

The likelihood formula is
\begin{align*}
    l(\boldsymbol{\theta};Y) &= \sum\limits_{M,q \in Q_M} \sum\limits_{[(i,j),(k,l)] \in \Phi_{M,q}} \log L(y_{ij},y_{kl};\boldsymbol{\theta}_{M,q}) \\
    &=\sum\limits_{M,q \in Q_M} l(\boldsymbol{\theta}_{M,q},Y)
\end{align*}
where $l(\boldsymbol{\theta}_{M,q},Y) = \sum\limits_{[(i,j),(k,l)] \in \Phi_{M,q}} \log L(y_{ij},y_{kl};\boldsymbol{\theta}_{M,q})$. Instead of finding the set of parameters that maximize $l(\boldsymbol{\theta};Y)$, we now find the set of parameters that maximize $l(\boldsymbol{\theta}_{M,q},Y) $. 

Let $s$ denote the index of optimization. For example, $\boldsymbol{\theta}_1=\boldsymbol{\theta}_{\phi_A,\{1,1\}}$, $\boldsymbol{\theta}_2=\boldsymbol{\theta}_{\phi_A,\{1,2\}}$,$\boldsymbol{\theta}_3=\boldsymbol{\theta}_{\phi_A,\{2,2\}}$, $\boldsymbol{\theta}_4=\boldsymbol{\theta}_{\phi_B,(1,\{1,1\})...}$. Let $\boldsymbol{\Theta}= [\boldsymbol{\theta}_1,\boldsymbol{\theta}_2,\boldsymbol{\theta}_3,\boldsymbol{\theta}_4,... ]$.
The asymptotic distribution of $\hat{\boldsymbol{\Theta}}$ is 
\begin{equation}
\sqrt{n} (\hat{\boldsymbol{\Theta}}-\boldsymbol{\Theta}) \overset{d}{\to}  MVN (0, A(\boldsymbol{\Theta})^{-1} B(\boldsymbol{\Theta}) A(\boldsymbol{\Theta})^{-1}),
\label{eqn:asymptotic_1}
\end{equation}
where $\displaystyle A(\boldsymbol{\Theta})= E\left[-\frac{\partial^2 \sum_s l_(\boldsymbol{\theta}_s,Y)}{\partial \boldsymbol{\Theta} \partial \boldsymbol{\Theta}'}\right]$ and $\displaystyle B(\boldsymbol{\Theta})= E\left[\frac{\partial \sum_s l(\boldsymbol{\theta}_s,Y)}{\partial \boldsymbol{\Theta}} \left(  \frac{\partial \sum_s l(\boldsymbol{\theta}_s,Y)}{\partial \boldsymbol{\Theta}} \right)' \right].$

Because $l_s(\boldsymbol{\theta}_s,Y)$ only involves $\boldsymbol{\theta}_s$, $A(\boldsymbol{\Theta})$ is a block-diagonal matrix with blocks 
\begin{center}
$\displaystyle A(\boldsymbol{\Theta})_{ss}= E\left[-\frac{\partial^2  l_s(\boldsymbol{\theta}_s,Y)}{\partial \boldsymbol{\theta}_s \partial \boldsymbol{\theta}_s'}\right] $,
\end{center}
and $B(\boldsymbol{\Theta})$ is a symmetric matrix where 
$\displaystyle B(\boldsymbol{\Theta})_{st}= E\left[\frac{\partial l_s(\boldsymbol{\theta}_s,Y)}{\partial \boldsymbol{\theta}_s} \left(  \frac{\partial l_t(\boldsymbol{\theta}_t,Y)}{\partial \boldsymbol{\theta}_t} \right)' \right]  $. \vspace{4mm}

In \citet{fieuws2006pairwise} and \citet{solomon2017pseudo},  independent observations are drawn from a multivariate distribution and longitudinal observations on the same individual are correlated. Since we deal with network data, we can not simply calculate the empirical version of $B(\boldsymbol{\Theta})$ by taking averages with $\hat{\boldsymbol{\Theta}}$ substituted. Therefore, we make the modification that observations used in maximizing $l_s(\boldsymbol{\theta}_s,Y)$ and $l_t(\boldsymbol{\theta}_t,Y)$ are distinct. Then $\hat{B}(\boldsymbol{\Theta})_{st}=\boldsymbol{0}$ and we can get $\hat{A}(\boldsymbol{\Theta})_{ss}$ and $\hat{B}(\boldsymbol{\Theta})_{ss}$ by taking averages with $\hat{\boldsymbol{\theta}_s}$ and $\hat{\boldsymbol{\theta}}_t$ substituted. 

The last step in getting $\hat{\boldsymbol{\beta}}$ and $SE(\hat{\boldsymbol{\beta}})$ is to take weighted averages of $\boldsymbol{\theta}_s \;\; \forall s$. Let $\hat{\boldsymbol{\theta}}=\boldsymbol{A} \hat{\boldsymbol{\Theta}}$, where $\boldsymbol{A}$ is the matrix that calculates the weighted averages, with weights proportional to the sample size used in each optimization. Then
\begin{equation}
    \sqrt{n} (\hat{\boldsymbol{\theta}}-\boldsymbol{\theta}) \overset{d}{\to}  MVN (0, \boldsymbol{A} \Sigma(\hat{\boldsymbol{\Theta}})\boldsymbol{A}^{'}),
\label{eqn:asymptotic_2}
\end{equation}
where $\Sigma(\hat{\boldsymbol{\Theta}})$ is the covariance matrix for $\hat{\boldsymbol{\Theta}}$ obtained by using Equation \eqref{eqn:asymptotic_1}.

\section{Definitions and notation}
In this section, we formally define the notation defined conceptually in the paper.
$Q_M$, which is the the set of block pairs/triplets for dyad configuration $M$ given $[B]$ (Section 4.1) is defined as:
\begin{itemize}
    \item $Q_{\sigma^2}=\{  (u,v): u, v \in[B] \}$
     \item $Q_{\phi_A}=\{  \{u,v\}: u, v\in [B]\}$
     \item  $Q_{\phi_B}=\{  (u,\{v,w \}): u, v, w \in [B] \}$
     \item  $Q_{\phi_C}=\{  (u,\{v,w \}): u, v, w \in [B]\}$
     \item  $Q_{\phi_D}=\{  (u,v,w): u, v, w \in[B] \}$
\end{itemize}
$\Phi_{M,q}$ is defined as:
\begin{itemize}
    \item $\Phi_{\sigma^2,(u,v)}= \{ [(i,j),(i,j)]: i, j\in [n], i \neq j, {g}_i=u,   {g}_j=v\} $
    \item $\Phi_{\phi_A,\{u,v\}}= \{ [(i,j),(j,i)]:i, j\in [n], i \neq j, {g}_i=u,   {g}_j=v \}$
    \item $\Phi_{\phi_B,(u,\{v,w\})}= \{ [(i,j),(i,k)]: i,j, k\in [n], i \neq j \neq k,{g}_i=u,   {g}_j=v,  {g}_k=w  \}$
    \item $\Phi_{\phi_C,(u,\{v,w\})}= \{ [(j,i),(k,i)]: i, j, k \in [n], i \neq j \neq k,{g}_i=u,   {g}_j=v,  {g}_k=w  \}$
    \item $\Phi_{\phi_D,(u,v,w)}= \{ [(i,j),(k,i)]:  i,j, k\in [n], i \neq j \neq k ,{g}_i=u,   {g}_j=v,  {g}_k=w \}$
\end{itemize} 
$\Phi_{M,i}$ is defined as:
\begin{itemize}
    \item $\Phi_{\sigma^2,i}= \{ [(i,j),(i,j)]: j\in [n], i \neq j \} \cup  \{ [(j,i),(j,i)]: j\in [n], i \neq j \}$
    \item $\Phi_{\phi_A,i}= \{ [(i,j),(j,i)]: j\in [n], i \neq j \}$
    \item $\Phi_{\phi_B,i}= \{ [(i,j),(i,k)]: j\in [n], k\in [n], i \neq j \neq k \}$
    \item $\Phi_{\phi_C,i}= \{ [(j,i),(k,i)]: j\in [n], k\in [n], i \neq j \neq k \}$
    \item $\Phi_{\phi_D,i}= \{ [(i,j),(k,i)]: j\in [n], k\in [n], i \neq j \neq k \}$
\end{itemize}

\bibliographystyle{abbrevnamed}
\bibliography{ref}

\end{document}